\documentclass[12pt]{article}
\input{epsf.sty}
\def\mydate{}
\def\ignore#1{{}}

\let\oldtheequation=\theequation
\def\doteqs#1{\setcounter{equation}{0}
            \def\theequation{{#1}.\oldtheequation}}
\newcounter{sxn}
\def\sx#1{\addtocounter{sxn}{1} \vskip 1.cm  \goodbreak
\noindent{\large\bf\leftline{\thesxn.~~#1}} \nobreak \vskip -.5cm}
\def\sxn#1{\sx{#1} \doteqs{\thesxn}}

\newcounter{axn}
\def\ax#1{\addtocounter{axn}{1} \bigskip\medskip\goodbreak \noindent{\large\bf
{\Alph{axn}.~~#1}} \nobreak \medskip}

\def\axn#1{\ax{#1} \doteqs{\Alph{axn}}}

\date{}

\newdimen\mybaselineskip
\renewcommand{\baselinestretch}{1.25}

\newcommand{\beeq}{\begin{equation}}
\newcommand{\eneq}{\end{equation}}
\newcommand{\beqn}{\begin{eqnarray}}
\newcommand{\eeqn}{\end{eqnarray}}

\def\dd{\partial}
\def\la{\raise.16ex\hbox{$\langle$}\lower.16ex\hbox{}  }
\def\ra{\, \raise.16ex\hbox{$\rangle$}\lower.16ex\hbox{} }
\def\go{\rightarrow}

\def\next{{~~~,~~~}}
\def\onehalf{ \hbox{${1\over 2}$} }

\def\ep{\epsilon}

\begin{document}
\thispagestyle{empty}

\baselineskip=12pt

{\small \noindent \mydate  \hfill OU-HET 402 /2002}

 {\small \hfill \bf hep-th/0201141}

\baselineskip=40pt plus 1pt minus 1pt

\vskip 3cm

\begin{center}

{\LARGE \bf Cosmology in the Einstein-Electroweak Theory
and Magnetic Fields}\\

\vspace{2.0cm}
\baselineskip=20pt plus 1pt minus 1pt

{\bf  Hiroki Emoto, Yutaka Hosotani and Takahiro Kubota}\\
\vspace{.1cm}
{\it Department of Physics, Osaka University,
Toyonaka, Osaka 560-0043, Japan}\\
\end{center}

\vskip 2.cm
\baselineskip=20pt plus 1pt minus 1pt

\begin{abstract}
In the $SU(2)_{L}\times U(1)_{Y}$ standard electroweak theory 
coupled with the Einstein gravity, 
new topological configurations naturally emerge, if the spatial 
section of the universe is globally a three-sphere
($S^3$) with  a small radius.  The $SU(2)_L$ gauge fields and
Higgs fields wrap the space nontrivially, residing at or near a 
local minimum of the potential.  As the universe expands, however, 
the shape of the potential rapidly changes and   
the local minimum eventually disappears.  The fields then start to roll
down  towards the absolute minimum.
In the absence of the $U(1)_Y$ gauge interaction the resulting space
is a homogeneous and isotropic $S^3$, but the $U(1)_Y$ gauge interaction
necessarily induces anisotropy  while preserving the homogeneity of the
space. 
Large magnetic fields are generically  produced
over a substantial period of the rolling-over transition.
The magnetic field configuration is characterized by the Hopf
map.
\end{abstract}

\centerline{PACS:  04.90+e, 11.15.Kc, 11.15.Ex, 12.15.-y}

\newpage


\newpage

\sxn{Introduction }

It has been well known that the standard electroweak theory 
does not admit classical lump solutions such as monopoles. 
There has been, however, unfailing interest in
non-perturbative  phenomena in the electroweak theory
such as baryon number violation by anomaly \cite{thooft} or sphalerons 
\cite{manton} and dumbbell solutions\cite{nambu,einhorn}. 
Such non-perturbative effects are expected to play a key role in 
particle physics phenomenology and cosmology.

Not so well investigated are cosmological configurations in the
electroweak theory, or more generally in non-Abelian gauge theory.  
Many years ago,  a classical exact solution in the $SU(2)$ Yang-Mills
theory coupled to  the Einstein gravity was found
\cite{jacob,hosotani1} which  describes a closed
Friedmann-Robertson-Walker universe with  large time-dependent
magnetic fields filling the space.  It was noticed that there
appears a natural map between the gauge group $SU(2)$ and the space $S^3$.
The extension to more general gauge group and spacetime has been
made.\cite{Molnar}  Dynamics of fermions are also
investigated.\cite{Gibbons} The analysis has been extended to a
semiclassical theory in the Euclidean signature in which solutions are
wormholes  describing quantum transitions to another
universe.\cite{hosoyaogura,Rey}

In the standard electroweak theory the gauge group is not $SU(2)$, 
but $SU(2)_L \times U(1)_Y$.  Furthermore the symmetry is
spontaneously broken to $U(1)_{EM}$ by  the Higgs fields having 
a nonvanishing expectation value.   We shall show that, in spite of 
these intricate features, the electroweak theory  leads to a new 
type of cosmological solutions.  The presence of the Higgs fields stabilizes 
topological configurations of the $SU(2)_L$ gauge fields when the size of
the universe is sufficiently small.  The $U(1)_Y$ gauge interaction 
deforms the map between the $SU(2)_L$ gauge group and the space, 
giving rise to anisotropy in the space.  The resulting space
is a deformed $S^3$ which is an anisotropic, but homogeneous compact
manifold.\cite{Barrow}  It may be recalled that such minimal distortion of
symmetry arises in sphaleron solutions as well; sphalerons are
spherically  symmetric in the $SU(2)_L$ theory, whereas they become  only
axially symmetric in the $SU(2)_L \times U(1)_Y$ theory \cite{manton}. 
In our case this deformation is related to the nontrivial Hopf map 
on $S^3$.

The gravity plays many vital roles in field theory.  Quantum effects,
for instance, lead to the Hawking effect around black holes
which shall prompt unification of thermal or statistical character with 
gravitational one.\cite{Hawking}   Novel effects emerge even in classical
theory. Due to attractive nature of gravitational force soliton-like
objects become possible even when such things are strictly forbidden 
in flat space.\cite{Galtsov,Bartnik, Breitenlohner}  In particular,
magnetic monopoles in pure Yang-Mills theory become possible in the
Einstein gravity with a negative cosmological constant.\cite{Hosotani2} 
Scalar field theory with a potential of the double well type admits cosmic
shells in asymptotically de Sitter space.\cite{Hosotani3, Daghigh,
Dymnikova}  In this paper we investigate another classical interplay of
gravity and field theory in the context of the cosmological evolution of
the universe.

The paper is organized as follows.   After giving general topological 
consideration of Higgs and gauge field configurations in Section 2, 
we give in Section 3 our ans{\" a}tze for the solutions to the equations 
of motion, whereby switching on and off the $U(1)_Y$ gauge interaction. 
The shape of the potential perceived by the Higgs 
and gauge fields is discussed in detail in Sect. 4. We demonstrate in
Section 5  that the Hopf mapping is  embodied  in our gauge field 
configuration, which makes it possible to distribute vector fields of 
constant magnitude all over the three-sphere. Time evolution of field
configurations  is evaluated  numerically in Section 6 for various input
parameters and  initial conditions. It is shown  in Section 7 that
$U(1)_{EM}$ magnetic field  survives the cosmological evolution for a
substantial period  of time  for  values of the  
parameters in a wide range.  Section 8 is devoted to summary 
and discussions.

\sxn{Topology in the Higgs and gauge fields}

In the present paper we discuss only the bosonic part of the
standard electroweak theory in the Einstein gravity whose action is given
by
\beqn
&&\hskip -1cm
I =\int d^4 x \sqrt{-g} ~
\bigg\{ \frac{1}{16\pi G} (R-2\Lambda )
-\frac{1}{4} F^{a}_{\mu \nu}F^{a\mu \nu}
-\frac{1}{4} G_{\mu \nu}G^{\mu \nu}   \cr
\noalign{\kern 8pt}
&&\hskip 3cm 
- (D_{\mu }\Phi)^{\dag} (D^{\mu}\Phi ) 
- \lambda \Big(\Phi ^{\dag}\Phi - \frac{v_0^2}{2} \Big)^{2} \bigg\} ~,   \cr
\noalign{\kern 8pt}
&&\hskip -1cm
D_{\mu }\Phi =(\partial_{\mu}- i \frac{g}{2} \, \tau^{a}{A_{\mu}^{a}}
-i \frac{g'}{2} B_{\mu})\Phi  ~, 
\label{action1}
\eeqn
where $F^{a}_{\mu \nu}$ and $G_{\mu \nu}$ denote the field strengths of
the $SU(2)_{L}$ and $U(1)_{Y}$ gauge fields $A^a_\mu$ and $B_\mu$, 
respectively.   $\Phi$ is a doublet Higgs field which develops a 
nonvanishing expectation value.  We employ the natural unit 
$\hbar=c=1$, and $G$ is the gravitational constant. 

We shall investigate  time evolution of classical field
configurations in a Robertson-Walker spacetime with a spatial section
$S^3$ or deformed $S^3$.  To understand why nontrivial topological
configurations appear in such spacetime, it is instructive to 
first examine  the topology of the electroweak theory on 
the fixed space $S^3$.

A  three-sphere $S^3$ with a radius $a$ is a hypersurface defined by
$z_1^2 + z_2^2 + z_3^2 + z_4^2 = a^2$ in the four-dimensional Euclidean 
space with Cartesian coordinates $\{ z_i \}$.  If the gauge interactions
are switched off, there appears a natural map between the Higgs vacuum
and $S^3$.  For later  convenience we  parameterize the Higgs field as
\beeq
\Phi=\frac{1}{\sqrt{2}} {\phi_2+i\phi_1 \choose \phi_4-i\phi_3}
\label{Higgs1x}
\eneq
where $\phi_i$'s are all real fields.  The manifold of the Higgs vacuum,
namely a manifold defined by minima of the Higgs potential, is 
$\phi_1^2+\phi_2^2+\phi_3^2+\phi_4^2 = v_0^2$, namely  $S^3$ in the
space of the Higgs fields.  There appears a natural map from the space
$S^3$ to the Higgs vacuum $S^3$;
\beeq
\phi_j = v_0 ~ y_j ~~,~~ y_j = \frac{z_j}{a} ~~.
\label{Higgs2x}
\eneq 
The Higgs fields wrap the space nontrivially.

Now let us switch on  the $SU(2)_L$ gauge interaction.  
By a gauge transformation,  the Higgs configuration (\ref{Higgs1x}) 
can be smoothly unwrapped. 
In fact by  a gauge transformation with a gauge potential
\beqn 
\Omega = y_4 + i \vec y \cdot \vec \tau \in SU(2) ,
\label{omega} 
\eeqn
the Higgs and gauge fields undergo, respectively,   transformation
\beqn
\Phi \quad &\go& \Phi' = \Omega^{-1} \Phi 
 = \Omega ^{-1} \frac{v_0}{\sqrt{2} } {y_2 + iy_1 \choose y_4 -i y_3}
= {0 \choose v_0/\sqrt{2} },  \cr
\noalign{\kern 8pt}
A=0  &\go& A' =  - \frac{i}{g} \, \Omega^{-1} d \Omega \not= 0 ~.
\label{gauge1}
\eeqn
The Higgs fields are brought into the standard form. This, however, does 
not imply that the effect of the wrapping  (\ref{Higgs1x}) of the Higgs 
field  has gone. 
The topological information of the wrapping is encoded in the
$SU(2)_L$ gauge   field (\ref{gauge1}) whose form is now far from
trivial.
There appears an energy barrier between (\ref{gauge1}) and the trivial 
configuration.  Less clear is the classical stability of the 
configuration. There is no topological index which guarantees 
the stability of the configuration.

The above consideration suggests that there are nontrivial topological 
configurations in the standard  electroweak theory. To be  realistic
we have to include the $U(1)_{Y}$ interaction and the dynamics of the 
curved space must be incorporated.   It shall turn out  that the
configuration (\ref{gauge1}) is stable only if  $gav_0$ is small enough,  
and that the Einstein equations dictate  $a$ to expand so that the
configuration can be stable only for a short period.  In the  following
sections we shall give thorough discussions both in the absence and
presence of the $U(1)_Y$ interaction.

\sxn{Configurations in the closed universe}

In the Einstein gravity, the existence of matter fields is 
 the source of distortion 
of the spacetime geometry. In the electroweak theory (\ref{action1}) 
 the energy-momentum tensor is given by 
\beqn
&&\hskip -1cm
T_{\mu\nu}
=  F^{a}_{\mu\rho} F_{\nu}^{a\rho}
- g_{\mu\nu} \frac{1}{4}F^{a}_{\rho\sigma}F^{a\rho\sigma}
+ G_{\mu\rho} G_{\nu}^{\rho}
- g_{\mu\nu} \frac{1}{4} G_{\rho\sigma} G^{\rho\sigma} \cr
\noalign{\kern 10pt}
&&
+(D_\mu \Phi )^{\dag}(D_\nu \Phi )+(D_{\nu} \Phi)^{\dag} (D_\mu \Phi) \cr
\noalign{\kern 10pt}
&&
- g_{\mu\nu} \bigg\{ (D_\rho \Phi)^\dag (D^{\rho}\Phi )
 + \lambda \Big(\Phi^\dag \Phi - {v_0^2 \over 2} \Big)^2 \bigg\} .
\label{EM-tensor1}
\eeqn
An ansatz for the metric must be consistently made with a configuration
of the gauge and Higgs fields.
To facilitate our discussions in switching on or off the $U(1)_Y$
interactions, it is most convenient to use differential
forms.  We in particular write the metric of the $S^3$ or $SU(2)$
manifold,  using the Maurer-Cartan forms.
Being equipped with them, we can easily go over  to  a deformed $S^3$
manifold which is necessary to describe the $SU(2)_L \times U(1)_{Y}$ 
electroweak theory in gravity.

\bigskip

\leftline{\bf 3.1 The Robertson-Walker spacetime $R^1 \times S^3$}

The Maurer-Cartan 1-forms, $\sigma_j$'s,  are expressed  in terms of
$\Omega $ in (\ref{omega}),  by 
\beeq 
\sigma^j \tau^j \equiv  -i\Omega^{-1} d\Omega ~~~,~~~
d\sigma^j = \ep^{jk\ell } \sigma^k \wedge \sigma^{\ell } ~~,
\label{form1}
\eneq
where $\tau_j$'s are Pauli matrices.  In terms of $y_j$
\beeq
\sigma_j = \ep^{jk\ell } \, y_k dy_{\ell} + y_4 dy_j - y_j dy_4 ~~.
\label{form2}
\eneq
The metric of a unit three-sphere is written as
\beeq
d {\Omega_3}^2 = \sigma^1 \otimes \sigma^1 +
 \sigma^2 \otimes \sigma^2 +\sigma^3 \otimes \sigma^3  
= {\sigma^a}_j {\sigma^a}_k \, dx^j  dx^k ~,
\label{metric1}
\eneq
which reduces in the spherical coordinates $(\chi, \theta, \phi)$ 
to the standard form 
$d {\Omega_3}^2 = d\chi^2 + \sin^2 \chi (d\theta^2+\sin^2\theta d\phi^2)$.

The metric and tetrads of the Robertson-Walker spacetime with a spatial
section $S^3$ is given by
\beqn
&&\hskip -1cm 
ds^2 = - N(t)^2 dt^2 + a(t)^2 d {\Omega_3}^2 , \cr
&&\hskip -1cm 
e ^{0}= N(t)dt \next e^j =a(t) \sigma^j \quad (j =1,2,3) ~.
\label{metric2}
\eeqn
The lapse function  $N(t)$ has been included for later convenience.

The curvature 2-forms in the tetrad basis are given by 
\beqn
&&\hskip -1cm 
\mathcal{R}_{jk} = A ~ e^j \wedge e^k 
\next A = \frac{1}{a^2}
 \bigg\{  1+  \Big( \frac{\dot a}{N} \Big)^2 \bigg\}  
\qquad (j, k=1,2,3), 
\cr
\noalign{\kern 10pt}
&&\hskip -1cm 
\mathcal{R}_{0j} = B~  e^0 \wedge e^j  
\next B =  - \frac{1}{aN} \frac{d}{dt} 
 \bigg( \frac{\dot a}{N} \bigg) \qquad (j=1,2,3) ~.
\label{curvature1}
\eeqn
The Ricci tensor $R_{ab}$ is diagonal.  
The Einstein tensor is
given by
\beeq
R_{ab} - {1\over 2} \eta_{ab} R 
= \pmatrix{3A \cr & -A + 2B \cr  && -A + 2B \cr &&& -A + 2B \cr} 
\label{Etensor1}
\eneq
where $\eta_{ab} = {\rm diag} (-1,1,1,1)$.
The space is homogeneous and isotropic.  

\bigskip

\leftline{\bf 3.2  In the $\theta_W = 0$ theory}

Suppose that the $U(1)_Y$ gauge interaction is absent, i.e.\ the 
weak mixing  angle vanishes; $\theta_W = \tan^{-1} (g'/g) =0$.  There
is a natural map between the space 
$S^3$ and $SU(2)_L$ gauge field configurations.  Following 
\ \cite{hosotani1}, we start from the following ansatz:
\beqn
&&\hskip -1cm
\Phi=\frac{1}{\sqrt{2}}{0 \choose v(t)}  \cr
\noalign{\kern 10pt}
&&\hskip -1cm
A=A^j_\mu \, \frac{\tau ^{j}}{2} \, d x^\mu
= \frac{1}{2g} f(t)~ \sigma^j \tau^j =-\frac{i}{2g} f(t)
 \Omega ^{-1}d\Omega \cr
\noalign{\kern 10pt}
&&\hskip -1cm
B= B_\mu dx^\mu = 0 ~~. 
\label{ansatz1}
\eeqn
Note that  $\Phi'$ and $A'$ in (\ref{gauge1}) corresponds to   
$v(t)=v_0$ and $f(t)=2$ in  (\ref{ansatz1}); 
the ansatz (\ref{ansatz1}) incorporates   non-trivial 
wrapping of configurations. 
We remark that each component of the gauge fields is,  
in spite of its simple appearance  in (\ref{ansatz1}),  endowed with 
nontrivial dependence in space, since  the orientation of the 
tetrad basis $\{ e^a \}$ varies in space. 
Nevertheless the configuration (\ref{ansatz1}) leads to a
self-consistent closed set of equations of motion. 

The $SU(2)_L$ field strength is
\beqn
F &=& dA+igA\wedge A \cr
\noalign{\kern 10pt}
&=& \left\{
{\dot f \over aN} ~  e^0 \wedge e^{\ell}
+ {f(2-f)\over 2 a^2} ~ \ep^{\ell mn} ~ e^m \wedge e^n \right\} 
{\tau^{\ell} \over 2g} ~.
\label{field1}
\eeqn
Note that $f(t)=2$ and $f(t)=0$ correspond to  pure gauge configurations.  
However, they are   physically distinct  and are
separated by  an energy-barrier when the Higgs fields are nonvanishing. 

Insertion of (\ref{ansatz1}) into (\ref{EM-tensor1}) gives 
energy-momentum tensors in the tetrad basis,  
$T_{ab} = {e_a}^\mu {e_b}^\nu \, T_{\mu\nu}$.
Off-diagonal components identically vanish.  Diagonal components are
\beqn
&&\hskip -1.2cm 
T_{00} = \frac{3}{2} \frac{\dot f^2}{ g^2 a^2 N^2} 
+\frac{\dot{v}^2}{2N^2} + V_{\theta_W=0}(v, f ; a) ~~, \cr
\noalign{\kern 10pt}
&&\hskip -1.2cm 
T_{11}= T_{22}= T_{33}=p(t),  
\label{EM-tensor2}
\eeqn
where the potential $V_{\theta_W=0}$ and the pressure $p(t)$ are  given, 
respectively,  by 
\beqn
V_{\theta_W=0}(v, f ; a) =  \frac{\lambda}{4} (v^2 - v_0^2 )^2
+\frac{3}{8}\frac{v^2 f^2}{a^2} 
+  \frac{3}{2} \frac{f^2(f-2)^2}{ g^2 a^4} ~~.
\label{potential1}
\eeqn
\beqn
p(t)=\frac{1}{2} \frac{\dot f^2}{ g^2 a^2 N^2} 
 +\frac{1}{2} \frac{f^2(f-2)^2}{ g^2 a^4} 
+\frac{\dot{v}^2}{2N^2}  - \frac{\lambda}{4} (v^2 - v_0^2 )^2
- \frac{1}{8}\frac{v^2 f^2}{a^2}.
\eeqn
Observe that $(v,f)=(v_0, 0)$ and $(v_0, 2)$ yield distinct
$T_{ab}$'s.  Each component of $T_{ab}$  does not depend on spatial 
coordinates $x^j$'s.
Further they preserve the rotational symmetry; $T_{0k}=0$ and
$T_{jk} =  p(t) \delta_{jk}$.  The pressure is the same in all spatial
directions.

The Einstein equation
\beqn
R_{ab} - \onehalf \eta_{ab} (R - 2 \Lambda) = 8\pi G T_{ab}, 
\eeqn
reduces to two equations.  
\beqn
&&\hskip -1cm
\frac{3}{a^2} \bigg\{  1+  \Big( \frac{\dot a}{N} \Big)^2 \bigg\}  
- \Lambda  = 8 \pi G T_{00}
\label{Ein1} \\
\noalign{\kern 8pt}
&&\hskip -1cm
 -\frac{2}{aN} \frac{d}{dt} \Big( \frac{\dot a}{N} \Big)
-\frac{1}{a^2} \bigg\{  1+  \Big( \frac{\dot a}{N} \Big)^2 \bigg\}   
+ \Lambda = 8 \pi G p(t)  ~~.
\label{Ein2}
\eeqn
The equations of motion of  the gauge  and Higgs fields are simplified
to  
\beqn
&&\hskip -1cm 
\frac{a}{N} \frac{d}{dt} \bigg( \frac{a}{N} \frac{df}{dt} \bigg)
+2f(f-1)(f-2)  +\frac{1}{4}(gva)^2 \, f=0 ~~, 
\label{YM1} \\
\noalign{\kern 8pt}
&&\hskip -1cm 
\frac{1}{a^3} \frac{d}{dt} \bigg(a^3 \frac{dv}{dt} \bigg)
+ \bigg\{ \lambda(v^2-v_0^2) +\frac{3}{4}\frac{f^2}{a^2} \bigg\}v=0 ~~.
\label{Higgs1}
\eeqn   
Not all of the four equations (\ref{Ein1}) $\sim$ (\ref{Higgs1}) are 
independent.  Eq.\ (\ref{Ein2}) follows from the other 
three.   The lapse function $N(t)$, which may be taken at will, is chosen
to be   $N(t)=1$ in the following discussions.  We have thus three
independent equations for three  unknown functions, i.e., $a(t)$, $f(t)$
and $v(t)$.   We comment that
Eqs. (\ref{Ein1})$\sim $ (\ref{Higgs1}) can also be 
obtained by first    inserting the ansatz into the action 
(\ref{action1}), and then by varying the action with respect to  
$N$, $a$, $f$, and $v$.

It is extremely   intriguing that the potential 
$V_{\theta_W=0}(v, f, a)$ in (\ref{potential1}) has a nontrivial
minimum in $(v, f)$ space in addition to the trivial one 
$(v, f)=(v_{0}, 0)$, when the scale factor $a$ is small enough.
(A more detailed account will be given in subsections 4.1 and 4.2.)  
The static configuration at the minimum in $(v, f)$ space 
would be a solution to (\ref{YM1}) and (\ref{Higgs1}), 
provided the time evolution of the scale factor $a$ could be frozen.  
As time develops, however, the scale factor $a$ necessarily  evolves 
subject to the Einstein equations  
and the shape of the potential accordingly changes. 
It is  interesting  to investigate 
the time development in the $(v, f)$ space starting from the 
local minimum by solving (\ref{Ein1}), (\ref{YM1}) and 
(\ref{Higgs1}).  
Before jumping into such enterprise, however, we have 
to refine the ansatz to include  the $U(1)_Y$ gauge
interaction. 

\bigskip
\leftline{\bf 3.3  In the $\theta_W \not= 0$ theory}

In our real world,  there is the $U(1)_Y$ gauge interaction whose presence
gives  an effect on the symmetry and structure of the universe.
We shall see that the resulting universe is homogeneous but anisotropic.

We need to generalize the ansatz (\ref{ansatz1}).  
The spatial component of the  $U(1)_Y$ gauge field necessarily 
picks one particular direction on $S^3$, giving
rise to  anisotropy.  This in turn affects and deforms the 
$SU(2)_L$ symmetry as well.
We fix the Higgs field in the standard form given in (\ref{ansatz1}),
employing  an $SU(2)_L$ gauge transformation.  With this choice the
$U(1)_Y$  gauge field $B$ must be proportional to $e^3$ and the asymmetry
in  the $SU(2)_L$ gauge fields must be aligned along  this
direction.   This is confirmed
a posteriori by computing energy-momentum tensors of the configuration.

The ansatz for the fields is given by 
\beqn
&&\hskip -1cm
\Phi=\frac{1}{\sqrt{2}}{0 \choose v(t)} ~,  \cr
\noalign{\kern 10pt}
&&\hskip -1cm
A = \frac{1}{2g} ~ \left\{  f_1(t) ( \sigma^1 \tau^1 + \sigma^2 \tau^2 )
 + f_3(t) \sigma^3 \tau^3 \right\}  ~, \cr
\noalign{\kern 10pt}
&&\hskip -1cm
B= h(t) \sigma^3  ~~. 
\label{ansatz2}
\eeqn
The resulting space is a deformed three-sphere.  The metric and tetrads of the
spacetime are
\beqn
&&\hskip -1cm 
ds^2 = - N(t)^2 dt^2 + 
a_1(t)^2 (\sigma^1 \otimes \sigma^1 + \sigma^2 \otimes \sigma^2 )
 + a_3(t)^2 \sigma^3 \otimes \sigma^3  ~, \cr
\noalign{\kern 10pt}
&&\hskip -1cm 
e ^{0}= Ndt \next e^1 =a_1 \sigma^1 \next
e^2 =a_1 \sigma^2 \next  e^3 =a_3 \sigma^3 ~~.
\label{metric3}
\eeqn
In  Appendix A the Riemann curvature  and Ricci tensors
are summarized for the more general metric
$ds^2 = - N^2 dt^2 + \sum_{j=1}^3 (a_j)^2 \sigma^j \otimes \sigma^j$.
It is shown  there that even in this general metric the Ricci tensors 
are diagonal, depending only on time $t$.  It gives an anisotropic, 
but homogeneous space.

Non-vanishing components of 
the energy-momentum tensors for the configuration (\ref{ansatz2}) are
\beqn
&&\hskip -1.5cm 
T_{00} =
\frac{1}{2g^2N^2} \bigg\{\frac{2\dot{f_1}^2}{a_1^2} 
  +\frac{\dot{f_3}^2}{a_3^2} \bigg\} 
+\frac{\dot{h}^2}{2N^2a_2^2}  +\frac{\dot{v}^2}{2N^2}  
   + V(v, f_1, f_3, h; a_1, a_3) ~,  \cr
\noalign{\kern 15pt}
&&\hskip -1.5cm
T_{11} = T_{22} \cr
\noalign{\kern 5pt}
&&\hskip -.8cm
=\frac{1}{2g^2N^2}\frac{\dot{f_3}^2}{a_3^2}
    +\frac{\dot{h}^2}{2N^2a_3^2}  +\frac{\dot{v}^2}{2N^2} 
 - \frac{\lambda}{4}(v^2-v_0^2)^2\cr
\noalign{\kern 10pt}
&&\hskip .5cm
-\frac{v^2}{8}\frac{(f_3-g'h)^2}{a_3^2} 
+\frac{1}{2g^2}\frac{(2f_3 - f_1^2)^2}{a_1^4}
+\frac{2h^2}{a_1^4} 
 ~, \cr
\noalign{\kern 10pt}
&&\hskip -1.5cm
T_{33} = \frac{1}{2g^2N^2} \bigg\{ \frac{2\dot{f_1}^2}{a_1^2}
 -\frac{\dot{f_3}^2}{a_3^2}  \bigg\}
- \frac{\dot{h}^2}{2N^2 a_3^2}
+\frac{\dot{v}^2}{2N^2} -\frac{\lambda}{4}(v^2-v_0^2)^2 \cr
\noalign{\kern 10pt}
&&\hskip -.5cm
-\frac{v^2}{8} \bigg\{ \frac{2f_1^2}{a_1^2}
      -\frac{(f_3-g'h)^2}{a_3^2}  \bigg\}
+\frac{1}{2g^2} \bigg\{ \frac{2f_1^2(2 - f_3)^2}{a_1^2 a_2^2}
    +\frac{(2f_2 - f_1^2)^2}{a_1^4}  \bigg\}  -\frac{2h^2}{a_1^4} ~~.
\label{EM-tensor3}
\eeqn
Here the potential in $T_{00}$ is given by
\beqn
&&\hskip -1.5cm 
V(v, f_1, f_3, h; a_1, a_3) 
= \frac{\lambda}{4}(v^2-v_0^2)^2 
+ \frac{v^2}{8} \bigg\{ \frac{2f_1^2}{a_1^2}
      +\frac{(f_3-g'h)^2}{a_3^2} \bigg\}  \cr
\noalign{\kern 10pt}
&&\hskip 1.5cm 
 +\frac{1}{2g^2} \bigg\{ \frac{2f_1^2(2 -f_3)^2}{a_1^2 a_3^2}
+\frac{(2f_3-f_1^2)^2}{a_1^4} \bigg\} +  \frac{2h^2}{a_1^4} ~~. 
\label{potential2}
\eeqn
In showing that $T_{ab}=0$ for $a\not= b$, the alignment of 
$B$ and $\Phi$  is crucial.  

The Einstein equations are 
\beqn
&&\hskip -1.5cm 
\frac{1}{N^2}\Bigl(\frac{\dot a_1^2}{a_1^2}
+\frac{2\dot{a_1}\dot{a_3}}{a_1 a_3}\Bigr)
+\frac{1}{a_1^2}\Bigl(4-\frac{a_3^2}{a_1^2}\Bigr)
- \Lambda = 8\pi G T_{00}  ~,  
  \label{Ein3}  \\ 
\noalign{\kern 10pt}
&&\hskip -1.5cm
\frac{1}{Na_1}\frac{d}{dt}\Big(\frac{\dot{a_1}}{N}\Big)
+\frac{1}{N a_3}\frac{d}{dt}\Big(\frac{\dot{a_3}}{N}\Big)
+\frac{\dot{a_1}\dot{a_3}}{N^2 a_1 a_3}+\frac{a_3^2}{a_1^4}
  -\Lambda = - 8\pi G T_{11}  ~, 
  \label{Ein4}  \\ 
\noalign{\kern 10pt}
&&\hskip -1.5cm
\frac{2}{Na_1} \frac{d}{dt} \Big( \frac{\dot{a_1}}{N} \Big)
+ \frac{\dot{a_1}^2}{N^2 a_1^2}
+\frac{4}{a_1^2} -\frac{3a_3^2}{a_1^4} -\Lambda = - 8\pi G T_{33} ~. 
  \label{Ein5}   
\eeqn
All off-diagonal components identically vanish.

Equations for the gauge fields and Higgs field are
\beqn
&&\hskip -1.cm
d({}^*F) -  ig(A\wedge {}^*F- {}^*F \wedge A)
  = - {}^*j^{SU(2)}  ~,  \cr
\noalign{\kern 5pt}
&&\hskip -1.cm
d({}^*dB) = - {}^*j^{U(1)} ~,  \cr
\noalign{\kern 5pt}
&&\hskip -1.cm
\frac{1}{\sqrt{-g}} D_\mu \Big\{
\sqrt{-g}g^{\mu\nu} D_\nu \Phi \Big\} 
=  \lambda \Big( 2\Phi^\dagger \Phi - v_0^2 \Big) \Phi ~,   
\label{matter1}
\eeqn
where ${}^*$ denotes  Hodge dual, and the currents are given by
\beeq
{}^*j^{SU(2)} =
\frac{\delta \mathcal{L}^{\rm Higgs}}{\delta A^{a\mu}}  
  \frac{\tau^a}{2} dx^\mu
\next 
{}^*j^{U(1)}
=\frac{\delta\mathcal{L}^{\rm Higgs}}{\delta B^{\mu}} dx^\mu  ~. 
\label{current1}
\eneq
It is easy to see  that $d({}^* A)=0$ and $d({}^*B) = 0$. Upon the
insertion of the ansatz, these equations are reduced to
\beqn
\frac{1}{N a_3} \frac{d}{dt} \Big(\frac{a_3}{N}\dot{f_1}\Big)
&=& - {1\over 2} g^2 a_1^2  \frac{\dd V}{\dd f_1} \cr
&=& - \bigg\{ {1\over a_3^2} (2- f_3)^2  
   -  \frac{1}{a_1^2}  (2f_3 - f_1^2) 
   + \frac{1}{4}(gv)^2 \bigg\} f_1 ~, 
  \label{YM3} \\
\frac{a_3}{N a_1^2}\frac{d}{dt}
\Bigl(\frac{a_1^2}{Na_3}\dot{f_3}\Bigr)
&=& -  g^2  a_3^2 \frac{\dd V}{\dd f_3} \cr
&=& - {2 a_3^2\over a_1^4} (2f_3 - f_1^2) 
   + \frac{2}{a_1^2} f_1^2(2 - f_3) 
  - \frac{1}{4} (gv )^2(f_3-g'h)  ~, 
  \label{YM4}  \\
\frac{a_3}{Na_1^2}\frac{d}{dt}\Bigl(\frac{a_1^2}{Na_3}\dot{h}\Bigr)
&=& -   a_3^2 \frac{\dd V}{\dd h} \cr
&=& - {4 a_3^2\over a_1^4} h 
+ \frac{1}{4} g'v^2 (f_3-g'h)  ~, 
  \label{U1}  \\
\frac{1}{Na_1^2a_3}\frac{d}{dt} \Bigl(\frac{a_1^2 a_3}{N}\dot{v}\Bigr)
&=& -  \frac{\dd V}{\dd v} \cr
&=& - \bigg[ \frac{1}{4} \bigg\{
\frac{2f_1^2}{a_1^2}+\frac{(f_3-g'h)^2}{a_3^2} \bigg\}
+\lambda(v^2-v_0^2)\bigg]v ~~. 
  \label{Higgs2}  
\end{eqnarray}
Inserting the ansatz into the action (\ref{action1}), one finds
\beqn
&&\hskip -2cm 
I = 2\pi ^{2} \cdot \frac{1}{16\pi G}\int dt a_{1}^{2}a_{3} \Bigg \{
\frac{4}{a_{1}}\frac{d}{dt}\left (\frac{\dot a _{1}}{N}\right )
+\frac{2}{a_{3}}\frac{d}{dt}\left (\frac{\dot a _{3}}{N}\right )
+\frac{2}{N}\left ( \frac{\dot a_{1}}{a_{1}}\right )^{2} 
+\frac{2}{N}\left ( \frac{\dot a_{1}}{a_{1}}\right ) \left (
\frac{\dot a_{3}}{a_{3}}\right )
\cr
\noalign{\kern 5pt}
&&\hskip 5cm 
+N\left ( \frac{8}{a_{1}^{2}}
 -\frac{2a_{3}^{2}}{a_{1}^{4}} \right ) -N \Lambda \Bigg \}
\cr
\noalign{\kern 5pt}
&&\hskip -1cm 
 +2\pi ^{2} \int dt Na_{1}^{2}a_{3}
\Bigg \{
\frac{1}{2g^{2} N^{2}}\left ( 
\frac{2 \dot f_{1}^{2}}{a_{1}^{2}}+\frac{ \dot f_{3}^{2}}{a_{3}^{2}}
\right ) 
+\frac{\dot h ^{2}}{2N^{2}a_{3}^{2}}
+\frac{\dot v ^{2}}{2N^{2}}
\Bigg \}
\cr
\noalign{\kern 5pt}
&&\hskip -1cm 
 -2\pi ^{2}  \int dt N a_{1}^{2}a_{3} V(v, f_{1}, f_{3}, h; a_{1}, 
 a_{3})   ~~.
\label{action2}
\end{eqnarray}
All of the equations, (\ref{Ein3})-(\ref{Ein5}) and 
(\ref{YM3})-(\ref{Higgs2}), have been obtained by inserting the ansatz 
into the equations of motion. They can   be 
obtained by first putting the ansatz into the action and then   
varying  the action  (\ref{action2})
with respect to $N, a_1, a_3, f_1, f_3, h$ and $v$.

\sxn{Potential in the fixed metric}

Before examining the time evolution of the field configurations,
it is most appropriate to understand the shape of the potential,  
supposing that the background metric is fixed.     
The emergence of a new local minimum
in the potential is a crucial ingredient  in our scenario in order 
to make it plausible to suppose that the universe once assumes  
 a topologically non-trivial field configuration.    We examine the
$\theta_W=0$ case first for which the location of the minima of the
potential can be  analytically determined, and then proceed to the
$\theta_W \not= 0$ case.

\bigskip
\leftline{\bf 4.1  In the $\theta_W=0$ theory}

In this case the potential $V_{\theta_W=0}$ in (\ref{potential1})
depends on the two variables $v (\ge 0)$ and $f$.  We write it in the form
\beqn
&&\hskip -1cm
V_{\theta_W=0} = 
\lambda v_0^4 \Bigg\{
 \frac{1}{4}  \bigg(  {v^2\over v_0^2} - 1 \bigg)^2  
+ {3\alpha\over8  \beta^2}  {v^2\over v_0^2} \, f^2
+ \frac{3 \alpha}{2 \beta^4}  f^2(f-2)^2  \Bigg\} ~~,  \cr
&&\alpha = {g^2\over \lambda} \next \beta = gav_0  ~~.
\label{potential3}
\eeqn
This shows that the shape of the potential depends on two dimensionless
parameters $\alpha$ and $\beta$.   In the standard model 
$\alpha = {\cal O}(1)$.
$\beta$ depends on the scale factor $a$.

\begin{figure}[tbh]\centering
\leavevmode 
\mbox{
\epsfxsize=8cm 
\epsfbox{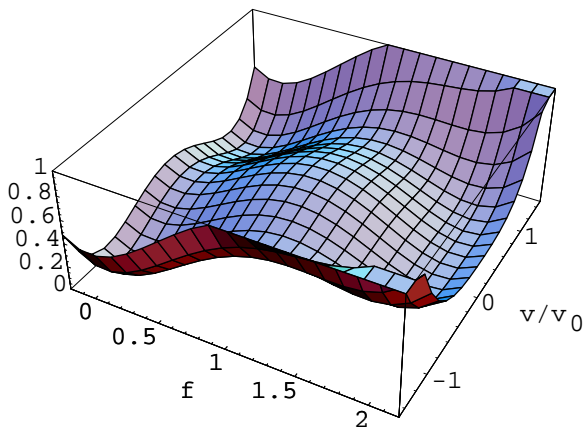}}   
\caption{SU(2) gauge-Higgs potential (\ref{potential3}) as a function of 
$f$ and $v/v_{0}$ ($\alpha=0.426$  and $\beta=1.14$)}
\label{fv-potential}
\end{figure}

If $\beta \gg 1$, the first term dominates over the rest
 in (\ref{potential3}) so that there appears only one minimum at
$v \sim v_0$ and $f=0$.  Less trivial is the case in which $\beta$ becomes
${\cal O}(1)$ or smaller.  The conditions for extrema are
\beqn
f  \bigg\{ f^2- 3f +2 +\frac{1}{8}
   \beta^2 {v^2\over v_0^2} \bigg\} =0   ~~, \cr
\noalign{\kern 10pt}
{v\over v_0} ~ \bigg\{
{v^2\over v_0^2} - 1 +\frac{3\alpha}{4\beta^2} ~f^2  \bigg\} =0  ~~.
\label{condition0}
\eeqn   
Nontrivial extrema appear for
\beeq
 \alpha <  {32\over 3} \next 
  \beta   <  \beta_c =
  \left(  2 \cdot {1+ {3\over 4} \alpha \over 1 - {3\over 32} \alpha}
   \right)^{1/2}     ~~.
\label{condition1}
\eneq
Let us define
\beqn
&&\hskip -1cm
 f_\pm 
=\frac{1}{2(1-\frac{3}{32}\alpha)} 
\left\{ 3 \pm \sqrt{ 9-\Bigl(1-\frac{3}{32}\alpha \Bigr)
             \Bigl(8+\frac{1}{2}\beta^2\Bigr) } ~  \right\}  ~,   
\label{fplusminus}
\\
\noalign{\kern 10pt}
&&\hskip -1cm
 v_\pm = v_0 \sqrt{ 1-\frac{3\alpha}{4\beta^2} f_\pm^2}  ~~.
\label{condition2}
\eeqn
A local minimum is located at
\beeq
(v, f) = \cases{
(v_+,  f_+) &for $\sqrt{3\alpha} < \beta < \beta_c$  ~, \cr
~(0, 2)&for $\beta < \sqrt{3\alpha}$  ~, \cr}
\label{condition3}
\eneq
The global minimum is always located at $(v, f) =(v_0, 0)$. 
For $\sqrt{3\alpha}/2  < \beta < \beta_c$, $(v_-, f_-)$  is a
saddle point.  The local minimum  is separated from the global minimum by
a barrier.  An example of the  potential is depicted in Figure
\ref{fv-potential}.

\begin{figure}[tbp]\centering
\leavevmode 
\mbox{
\epsfxsize=8cm 
\epsfbox{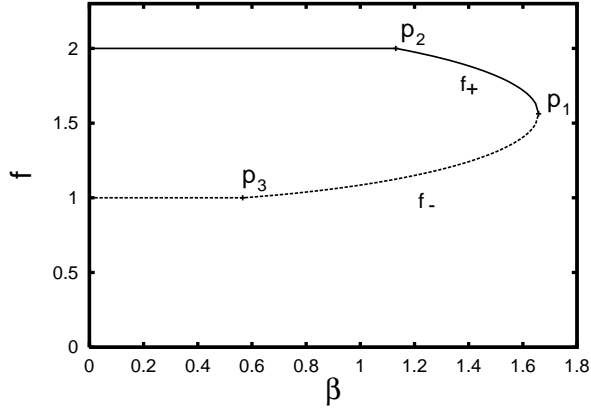}}   
\caption{Location of the extrema $f_{\pm }$  (eq. (\ref{fplusminus})) 
as a function of $\beta$
for $\alpha=0.426$.  $f_+$ and 
$f_-$ correspond to the local minimum and saddle point, respectively.
P$_1$, P$_2$, and P$_3$ have $\beta=\beta_c=1.658, \sqrt{3\alpha}=1.131,$ and
$\sqrt{3
\alpha}/2=0.5655$
respectively.}
\label{f-beta}
\end{figure}

\begin{figure}[tbp]\centering
\leavevmode
\mbox{
\epsfxsize=8cm 
\epsfbox{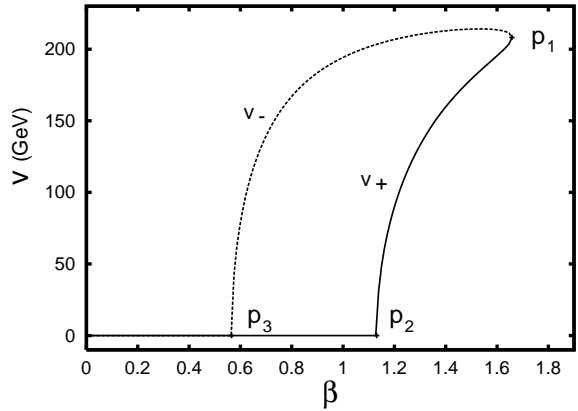}}   
\caption{Location of the extrema $v_{\pm }$ (eq. (\ref{condition2})) 
 as a function of $\beta$.  $v_+$ and
$v_-$ correspond to the local minimum and  saddle point, respectively.
At P$_3$, the local minimum and saddle point merge.}
\label{v-beta}
\end{figure}

The location of the local minimum varies as $\beta$.  With a given 
value of $\alpha$, $f_+(\beta)$ monotonically decreases from 2 to
${3\over 2} ( 1 - {3\over 32} \alpha)^{-1}$ as $\beta$ varies from 
$\sqrt{3\alpha}$ to $\beta_c$.  The $\beta$-dependence of the local
minimum $(v,f)$ is depicted in Figures \ref{f-beta} and \ref{v-beta}.
Here and hereafter we use values 
\beqn
v_0 \approx 246 {\rm GeV}, \qquad g\approx  0.653,  \qquad 
g' \approx  0.358
\label{suuji1}
\eeqn 
as the electroweak parameters.  As to $\lambda $ we put 
\beqn
\lambda \approx  1
\label{suuji2} 
\eeqn
as a generic value.

In the above considerations, our analysis of Eq. (\ref{condition0}) 
has been restricted to the case $\alpha <32/3$ in accordance with our 
choice of the parameters,  (\ref{suuji1}) and (\ref{suuji2}). For 
completeness we give a short remark as to the case $\alpha > 32/3$. 
Eq. (\ref{condition0}) gives us  obviously solutions $(v, f)=(v_{0}, 0)$, 
and $(0, 2)$ in this case as well. 
Apparently, $(v_{0}, 0)$ is the absolute minimum of the potential . 
On the other hand $(0, 2)$ is a local minimum for $\beta < \sqrt{3\alpha }$ 
and is a saddle point for $\beta > \sqrt{3\alpha}$. 
In addition to these solutions, there exists 
another solution $(v_{-}, f_{-})$, provided that 
$\sqrt{3\alpha }/2 < \beta < \sqrt{3\alpha }$. This solution 
 is always a saddle point. 
In any way,  we do not consider the case $\alpha > 32/3$ 
hereafter.

\bigskip
\leftline{\bf 4.2 In the $\theta_W \not= 0$ theory}

The potential $V$ defined in (\ref{potential2}) may be split into three 
terms according to the power behavior with respect to the scales $a_1$ and 
$a_3$;
\beqn
&&\hskip -1.cm 
V(v, f_1, f_3, h; a_1, a_3)  = V_0 + V_2 + V_4 \cr
\noalign{\kern 10pt}
&&\hskip 0.cm 
V_0 = \frac{\lambda}{4}(v^2-v_0^2)^2 \cr
\noalign{\kern 10pt}
&&\hskip 0.cm V_2 =  \frac{v^2}{8} \bigg\{ \frac{2f_1^2}{a_1^2}
      +\frac{(f_3-g'h)^2}{a_3^2} \bigg\}  \cr
\noalign{\kern 10pt}
&&\hskip 0.cm 
V_4 = \frac{1}{2g^2} \bigg\{ \frac{2f_1^2(2 -f_3)^2}{a_1^2 a_3^2}
+\frac{(2f_3-f_1^2)^2}{a_1^4} \bigg\} +  \frac{2h^2}{a_1^4} ~~. 
\label{potential4}
\eeqn
It is a function of  four variables, $v, f_1, f_3$ and $h$.  It  also
depends on the values of the  two scale factors $a_1$ and $a_3$.  As we
shall see below, the difference between $a_1$ and $a_3$ remains relatively
small in the  cosmological evolution.  
The global minimum is located at
$(v, f_1 , f_3 , h) = (v_0, 0, 0, 0)$.

As in the $\theta_W = 0$ case there appears a new local minimum 
when $ga_1 v_0$ and $ga_3 v_0$ are small enough.  
To pin down the location of the local minimum we again utilize the 
stationary conditions (\ref{YM4}) - (\ref{Higgs2}), ignoring the time 
dependence.  For $v \not= 0$ we have
\beqn
&&\hskip -1.cm
g'h = - {g'^2\over 2 g^2} \bigg\{ 2 f_3 - f_1^2 
 + {a_1^2\over a_3^3} f_1^2 (f_3 - 2) \bigg\}  ~,  
   \label{condition4}  \\
\noalign{\kern 10pt}
&&\hskip -1.cm
v^2 = v_0^2 - {1\over 4\lambda}
\Bigg\{ \frac{2f_1^2}{a_1^2}+\frac{(f_3-g'h)^2}{a_3^2} \Bigg\} ~~.
\label{condition5}
\eeqn
Insertion of (\ref{condition4}) and (\ref{condition5}) into
(\ref{potential4}) yields a  potential $\hat V$ as a function of $f_1$ and
$f_3$.  We look for  extrema of $\hat V$ under  the condition that the
right hand side of (\ref{condition5}) be positive.

The location of the new local minimum is not altered so much by
the presence of the $U(1)_Y$ gauge interaction.  As can be seen from 
(\ref{condition4}), the value of $g'h$ is very small for $f_1, f_3 \sim 2$.
However, this does not necessarily mean that the $U(1)_Y$ gauge 
interaction is unimportant.  In the course of expansion of the universe, 
 the local minimum disappears.
One would then ask  if the fields roll down towards the global minimum.
We shall see in Section 6 that in a wide range of the parameters in the
theory field configurations never reach  the global minimum.
As $a_1$ and $a_3$ become large,  the $V_4$ part of the potential $V$
becomes irrelevant. The relevant part of the potential $V_0 + V_2$ has a
flat direction along $v=v_0, f_1=f_3 - g'h=0$.  Neither $f_3$ nor $h$
approaches zero.  In such cases the $U(1)_{\rm EM}$ fields play an 
important  role in a substantial period of the expansion of the universe.

\sxn{$U(1)_Y$ gauge fields and the Hopf map}

As explained in Section 3 the presence of the $U(1)_Y$ gauge 
interaction alters the symmetry of the space.  The $U(1)_Y$ 
field strengths, both electric and magnetic, pick a preferred direction
at each space point, thus breaking the isotropy of the space.  
The homogeneity of the space is more subtle,  depending on the 
configuration.  We have found
that the configuration (\ref{ansatz2}) breaks the isotropy of the space,
but maintains its homogeneity.  How can it be possible to have  
nonvanishing $U(1)_Y$ field strengths  all over the compact space 
without spoiling  the homogeneity of the space?

To address the issue more precisely, we recall the $U(1)_Y$ field
strengths are given by
\beeq
 dB  = {\dot h\over a_3} \, e^0 \wedge e^3
 + {2h\over a_1^2} \, e^1 \wedge e^2
\label{Fstrength1}
\eneq
Both the electric and magnetic fields point in the $e^3$-direction at each 
point $\vec x$.  The direction varies in space as $\sigma_3(\vec x)$ is
$\vec x$-dependent.  The magnitudes of the fields, 
however, are independent of $\vec x$.  In other words we 
have a vector field $\vec K (\vec x)$
defined over an entire  compact space topologically equivalent to $S^3$. 
One may wonder what is ensuring such vector configuration on $S^3$.

It would be helpful to contrast the situation with
  a two-dimensional vector field on $S^2$.  Placing a vector
field 
$\vec K (\vec x)$ with $|\vec K|=
{\rm constant}$ necessarily induces  sources/sinks or vortices where
the vector field is ill defined.  It is thus impossible 
to have a constant two-dimensional vector field 
on $S^2$ without introducing  singularities.  In three dimensions,
however, such a vector field can be smoothly defined on $S^3$.  With the
condition 
$|\vec K|={\rm constant}$ the vector field defines a map from 
the space $S^3$ to
the $\vec K$ space $S^2$.  There exists a nontrivial map 
called  the Hopf map in mathematical literatures \cite{nakahara}.  
It is straightforward to check that 
$K  = k_0 \sigma^3$ is a Hopf map.

A  Hopf map is realized in electromagnetic $U(1)_{\rm EM}$ field 
as well.
\ignore{ 
The $U(1)$  field introduces the anisotropy, while maintaining 
the homogeneity.The resultant spacetime is a variant of 
the Robertson-Walker space whose
spatial section is a deformed $S^3$.  
}
In the standard electroweak theory
the nonvanishing Higgs field breaks the symmetry 
$SU(2)_{L} \times U(1)_{Y}$ down to
the electromagnetic $U(1)_{\rm EM}$.  Let us  define
\beqn
A_{\rm EM} &=& {1\over \sqrt{g^2 + g'^2}} 
\left\{ g h + g' {f_3\over g} \right\} \, \sigma^3
\equiv h_{\rm EM}  \sigma^3  , \cr
\noalign{\kern 20pt}
Z~ &=& {1\over \sqrt{g^2 + g'^2}} 
\left\{ -g' h + f_3 \right\} \, \sigma^3
\equiv h_{\rm Z}  \sigma^3  ~~, 
\label{EMZfield}
\eeqn
where 
$h_{\rm EM}$ and $h_{\rm Z}$ correspond to the electromagnetic  and
$Z$-field, respectively.  The electromagnetic field strengths are
\beeq
F_{\rm EM} = {\dot h_{\rm EM}\over a_3}  e^0 \wedge e^3
+  {2 h_{\rm EM}\over a_1^2 }  e^1 \wedge e^2  ~~.
\label{Fstrength2}
\eneq
\ignore{
The potential $V_2$ in (\ref{potential4}) depends on only $f_1$ and $h_Z$.
W- and Z-bosons become massive, while photons remain massless.  
}
As we see above, both electric and magnetic 
 fields point in the $e^3$ direction.  
Their magnitude
depends on time, but is independent of space position.  These vector fields
thus realize a nontrivial Hopf map.

The electromagnetic current one-form is connected to the gauge field 
by $d({}^* d A_{\rm EM}) = - {}^* j_{\rm EM}$;
\beeq
j_{\rm EM} =  \frac{a_3}{a_1^4} \left\{  {a_1^2\over Na_3} \frac{d}{dt} 
   \bigg( {a_1^2 \over Na_3} \, \dot h_{\rm EM} \bigg)
 + 4 h_{\rm EM} \right\} \, e^3 ~~.
\label{current2}
\eneq
Making use of   Eqs.\ (\ref{YM4}) and (\ref{U1})  or 
\beeq
\frac{a_1^2}{Na_3}\frac{d}{dt}
\bigg( \frac{a_1^2}{Na_3} \, \dot h_{\rm EM} \bigg)
= - 4 h_{\rm EM} 
+ {2 \sin \theta_W\over g} \, f_1^2 \, 
\bigg\{ 1 + \frac{a_1^2}{a_3^2} (2 - f_3) \bigg\} 
\label{EMfield1}
\eneq
 in (\ref{current2}), one finds
\beeq
j_{\rm EM}=
{2 \sin \theta_W\over g a_1^2 a_3} \, f_1^2 \, 
\bigg\{ 2 + \frac{a_3^2}{a_1^2} - f_3 \bigg\}e^3  ~~.
\eneq
The magnitude of the current is space-position independent. The current
also realizes a non-trivial  Hopf map.   

\ignore{
In the next section we shall see that in the cosmological evolution
$f_1$ and $h_{\rm Z}$ approach zero, but $h_{\rm EM}$ does not.
The scaled current $a_1^2 a_3 j_{\rm EM}^3$ quickly becomes small as $f_1$,
whereas the scaled magnetic field $a_1^2 F_{\rm EM}^{12}$ remains
nonvanishing.
}

\sxn{Cosmological evolution }

We have seen in the preceding sections that a nontrivial local
minimum of the potential appears when the size of the universe is 
sufficiently small.   In such a case we can expect  interesting 
phenomena in the history of the universe. 
Suppose that our system starts from the local minimum at some stage 
of the universe.  
The configuration at the minimum  cannot be
stationary, however.  The Einstein equations dictate that the universe expand.
We need to solve the Einstein and field equations simultaneously
to find precisely how the configuration evolves.  
Without going into detailed calculation, however, we may surmise 
the followings. As the universe expands, the barrier separating the local 
minimum from the global minimum disappears.
If the expansion rate of the universe is slow enough, the field 
configuration  could reach the global minimum or its vicinity  within
finite time. If the  universe  expands very fast, however, 
the potential $V$ becomes so flat along the flat direction 
of $V_0 + V_2$, the field configuration would never reach the global
minimum.

As it turns out, interesting phenomena take place  when a non-vanishing  
cosmological constant $\Lambda$ drives the universe to reasonably
fast inflation. We are not going to ask the origin of inflation or  the
source of $\Lambda $.  Instead we limit ourselves to ask how large the
cosmological constant should be for an initial nontrivial  field
configuration to lead to observable effects.  We   suppose that at an
initial time $t_0$ the sizes of the universe, $a_1(t_{0})$ 
and $a_3(t_{0})$, are   small.
To get an  idea of the magnitude of $\Lambda $ 
for a nontrivial configuration to exist, we look at one of the Einstein
equations (\ref{Ein3}), supposing $a_1 \approx  a_3 (\equiv a)$.  
In the $N=1$ gauge  it reads
\beeq
3 \left(  \frac{\dot a^2}{a^2} + \frac{1}{a^2} \right) 
- \Lambda  \approx  8\pi G T_{00} ~~.
\label{Ein6}
\eneq
In the $\theta_W=0$ case the condition 
$ga(t_0)v_0\stackrel{\displaystyle <}{\raisebox{-1ex}{$\sim$}}\beta_c$
$\sim 1.658$ must be satisfied to have a nontrivial local
minimum as discussed in Section 4.1 with $g \sim 0.653$.
In $\theta_W \not = 0$ theory   the condition is  modified.
A little numerical computation shows that the condition
for the existence of a nontrivial local minimum in the potential
remains almost unaltered.  $T_{00}$ is  either $O(v_0^4)$ or $O(a^{-4})$
and the right hand side of  (\ref{Ein6}) is much smaller than $a^{-2}$ for
 $ga v_0 \stackrel{\displaystyle < }{\raisebox{-1ex}{$\sim$}}1.658$. 
In other words   
\beeq
a(t)= \sqrt{\frac{3}{\Lambda}} \cosh{\sqrt{\frac{\Lambda}{3} } \, (t-t_1)}
\label{scalefactor1}
\eneq
at $t \sim t_0$.   As $a(t)\stackrel{\displaystyle >
}{\raisebox{-1ex}{$\sim$}} (3/\Lambda)^{1/2}$, the cosmological constant
must be larger than
$3 (g v_0 /\beta_c)^2$ for a nontrivial configuration to exist.
It is therefore reasonable  to discriminate 
the following three cases:

{\bf Case I:}
$\Lambda \gg (gv_{0})^{2}$  \qquad [$\rho \gg (10^{11}\; {\rm GeV})^{4}$], 
          
{\bf Case II:}
$\Lambda \sim (gv_{0})^{2}$ \qquad [$\rho \sim (10^{11}\; {\rm GeV})^{4}$],
           
{\bf Case III:}
$\Lambda \ll (gv_{0})^{2}$  \qquad [$\rho \ll (10^{11}\; {\rm GeV})^{4}$].

\noindent 
Here $\rho $ is the energy density corresponding   to 
$\Lambda=8\pi G \rho$.  The GUT energy scale $\rho \sim (10^{15} {\rm
GeV})^{4}$ corresponds to  Case I and the electroweak scale $\rho \sim
(v_{0})^{4}$ to Case III. 
 Cases I and II both encompass those of  nontrivial local minimum being 
present, while Case III does not.

The  behavior of the various fields can be  understood in general terms. 
 The field $f_1$, $h_Z$ and $v$ share 
common behavior; they approach
the values at the global minimum of the potential $V$.  Take $f_1$, as an
example.  Eq.\ (\ref{YM3}) becomes 
\beeq
\ddot f_1 + {\dot a_3\over a_3} \dot f_1 
  + {1\over 4 }{a_1^2\over a_3^2} g^2 v^2 f_1
= - \left\{ {1\over a_3^2}(2- f_3)^2 +{1\over a_1^2}( 2f_3 - f_1^2) \right\}
\, f_1 .
\label{YM5}
\eneq
As $a_j$ becomes large, the right hand side becomes negligibly small. 
The second term of the left hand side of (\ref{YM5})  may be regarded as 
a  friction term.   In the inflationary phase $\dot a_j/a_j \sim
(\Lambda/3)^{1/2}$.  As justified a posteriori, $a_1 \sim a_3(\equiv a)$.
Eq.\ (\ref{YM5}) is approximated by
\beeq
 \ddot f_1 + \sqrt{{\Lambda\over 3}} \dot f_1 
  + {1\over 4 }  g^2 v^2 f_1 = 0 
\label{YM6}
\eneq
whose solution is given by
\beqn
&&\hskip -1cm
f_1 \sim f_1(t_1) \, e^{-c (\Lambda/3)^{1/2}(t-t_1)}
\sim f_1(t_1) \bigg[ {a(t_1)\over a(t) } \bigg]^c  ~~, \cr
\noalign{\kern 10pt}
&&\hskip 0cm
c = {1\over 2} \Bigg\{ 1 - 
  \bigg(1 - {3g^2 v_0^2 \over \Lambda} \bigg)^{1/2} \Bigg\} ~~.
\label{estimate1}
\eeqn
$f_1$ eventually approaches zero.

The situation is qualitatively  different  for $h_{\rm EM}$.
Eq.\ (\ref{EMfield1}) reads
\beeq
\ddot h_{\rm EM}  + 
 \bigg( {2 \dot a_1\over a_1} - {\dot a_3\over a_3} \bigg) 
 \dot h_{\rm EM} 
={a_3^2 \over a_1^4} \left\{  - 4 h_{\rm EM} 
+ {2 \sin \theta_W\over g} \, f_1^2 \, 
\bigg[ 1 + \frac{a_1^2}{a_3^2} (2 - f_3) \bigg] \right\} .  
\label{EMfield2}
\eneq
which is approximated by
\beeq
\ddot h_{\rm EM}  +  \sqrt{{\Lambda\over 3}}  \dot h_{\rm EM} =0 ~.
\label{EMfield3}
\eneq
The solution is
\beeq
h_{\rm EM}(t) 
= h_{\rm EM}^\infty + c' e^{-\sqrt{ \Lambda/3 } \cdot t}
=  h_{\rm EM}^\infty + {c'' \over a(t)} ~.
\label{estimate2}
\eneq
$h_{\rm EM}$ approaches an asymptotic value $h_{\rm EM}^\infty$
whose magnitude depends on the initial conditions.

In the following we shall solve 
Eqs. (\ref{Ein4}), (\ref{Ein5}), (\ref{YM3}) -- (\ref{Higgs2})
numerically.   We   use the numerical values given in (\ref{suuji1}) and
(\ref{suuji2}) for $g$, $g'$, $v_{0}$  and $\lambda $. 
For convenience we set 
$\dot f_{1}=\dot f_{3}=\dot h=\dot v=0$,  
 $a_{1}=a_{3}$, and $\dot a_{1}=\dot a_{3}$  at $t=t_0$.
 We start from a field configuration at or near the local minimum of the 
potential (\ref{potential2}). 
Initial values for $a_1=a_3$,
$f_1$, $f_3$, $h$, and $v$ are  chosen
to be consistent with Eq.\ (\ref{Ein3}) with a given $\Lambda$.
It is satisfied at $t=t_{0}$  if the potential           
on the right hand side of (\ref{Ein3})  and the scalar           
curvature and cosmological constant  on the left are balanced at       
the local minimum. The initial value for  $\dot a_{1}=\dot a_{3}$ 
is determined  consistently.   We have checked that Eq.\ (\ref{Ein3}) is
 satisfied and $a_1 \sim a_3$ in the subsequent evolution.

\bigskip

\leftline{\bf Case I. $\Lambda \gg (g v_0)^2$ ,  
$\rho \gg (10^{11}  \, {\rm GeV})^{4}$  }

\begin{figure}[htb]\centering
\leavevmode
\mbox{
\epsfxsize=8cm 
\epsfbox{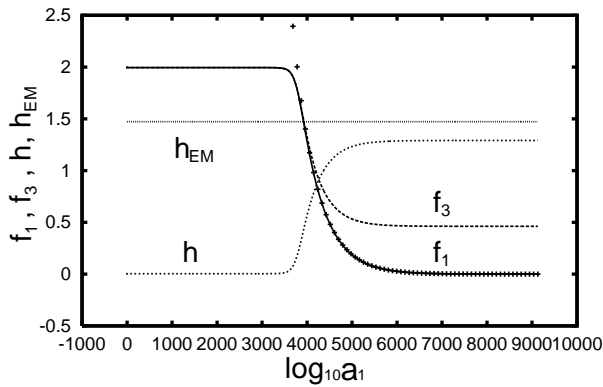}}   
\caption{The evolution of $f_1, f_3, h, h_{\rm EM}$ when
the configuration started from the local minimum of the potential.
In this example, $\Lambda=1.0 \times  10^7 $ GeV$^2$
and $a_{1}(t_0)= a_{3}(t_{0})= \sqrt{3/\Lambda}$.
In the transition region $f_1$ is well approximated by 
(\ref{estimate1})  with $c=0.001935$.}
\label{fh-largeL}
\end{figure}

This includes the GUT inflation where $\rho \sim (10^{15} \,{\rm GeV})^4$.
The universe  expands so fast that the fields can change only slowly.
It takes long time before the fields start to significantly evolve
to their asymptotic values.  The approach to the 
asymptotic values is governed by (\ref{estimate1}) and (\ref{estimate2}).
As $(gv_0)^2/\Lambda \ll 1$, 
\beeq
c = {3 g^2 v_0^2 \over 4 \Lambda} 
\label{estimate3}
\eneq
in (\ref{estimate1}).  One example is shown in Fig. \ref{fh-largeL}.

\ignore{
  For  $\Lambda = 1.7 \times 10^{10}\,$GeV$^2$, 
$a_j(t_0)=9 \times 10^{-3}\,$GeV$^{-1}$, we found that the local minimum is located
at $f_1 = f_3 = 1.958$, $h = 0.03283$, and $v = 158.5\,$GeV.  After
60 ten-fold growth of $a_j$, $\delta f_1= -1.074\times 10^{-4}$, 
$\delta f_3= -1.072\times 10^{-4}$,
$\delta h= 8.866 \times 10^{-5}$ and $\delta v = 0.04531$.
}

\begin{figure}[htb]\centering
\leavevmode
\mbox{
\epsfxsize=8cm 
\epsfbox{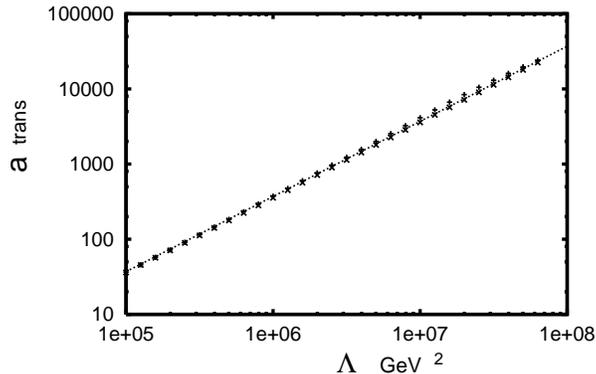}}   
\caption{For large $\Lambda$ the transition scale $a_{\rm trans}$
linearly depends on $\Lambda$.  In the plot $a_{\rm trans}$ is
defined by $f_1=1$ at $a=a_{\rm trans}$.  Points $+$ 
correspond to the evolution starting from the local minimum, while
 points $\times$  correspond to the evolution starting from the 
configuration $h=2$.  Little difference is seen.  The line
$a_{\rm trans} = 0.00037  \Lambda$ is drawn for
visual guide.
}
\label{trans}
\end{figure}

Let us denote by $a_{\rm trans}$ the scale factor $a$ at which  
the transition in the fields takes place.  The value of $a_{\rm trans}$
depends on $\Lambda$.  We have explored it numerically up to 
$\Lambda = 10^8\,$GeV$^2$
to find that $a_{\rm trans}$ is proportional to $\Lambda$;
\beeq
a_{\rm trans} \sim 1530 \cdot {\Lambda\over (gv_0)^3} ~~.
\label{transition}
\eneq
See Fig. \ref{trans}. It is of great interest to know 
why (\ref{transition}) holds.  It is also confirmed that the asymptotic
value for $h$ or $h_{\rm EM}$ depends on the initial value $h(t_0)$,
but depends little on $\Lambda$.

\noindent

\bigskip

\leftline{\bf Case II. $\Lambda \sim (g v_0)^2$ , 
  $\rho \sim (10^{11} \, {\rm GeV})^{4}$}

This is the most interesting case.    Suppose that the initial
configuration is at or very close to the local minimum of the potential,
and the corresponding $\beta$ in (\ref{potential3}) is well below 
the critical value $\beta_c$.  In this case
the field configuration  stays near the minimum for a while before
starting to roll down the hill of the potential.  If the
field configuration is away from the minimum, the fields quickly 
start to roll down toward the asymptotic values.
In case the initial $\beta$ is close to the critical value $\beta_c$,
the fields quickly undergo transition, irrespective of whether
the initial configuration is near or away from the local minimum. 

The approach to the asymptotic values of the fields is governed by
(\ref{estimate1}) again.
$f_1$  approaches zero in two to five ten-fold growth of
$a_j$, depending on the value of $\Lambda/g^2 v_0^2$. 
The way of approaching zero  depends on the initial values.
We illustrate it by taking the following two examples:

{\bf Case IIa}: $a_{1}(t_{0})=a_{3}(t_{0})=9.0 \times 10^{-3} 
 {\rm GeV}^{-1}$, 
$\Lambda =1.0 \times 10^{5}$ GeV$^2$, $\beta(t_0)= 1.446$

{\bf Case IIb}: 
$a_{1}(t_{0})=a_{3}(t_{0})=6.0 \times 10^{-3} {\rm   GeV}^{-1}$, 
$\Lambda =1.0 \times 10^{5}$ GeV$^2$, $\beta(t_0)= 0.964$.

\begin{figure}[tbh]\centering
\leavevmode 
\mbox{
\epsfxsize=8cm 
\epsfbox{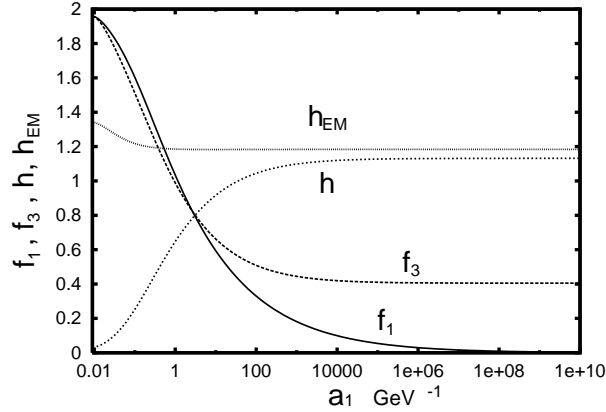}}   
\caption{The evolution of $f_1, f_3, h, h_{\rm EM}$ when
the configuration started from the local minimum of the potential.
In this example, $\Lambda=1.0 \times  10^5 $ GeV$^2$
and $a_{1}(t_0)= a_{3}(t_{0})= 9 \times 10^{-3} {\rm GeV}^{-1}$.  
}
\label{fh-evolution1}
\end{figure}

\begin{figure}[tbh]\centering
\leavevmode 
\mbox{
\epsfxsize=8cm 
\epsfbox{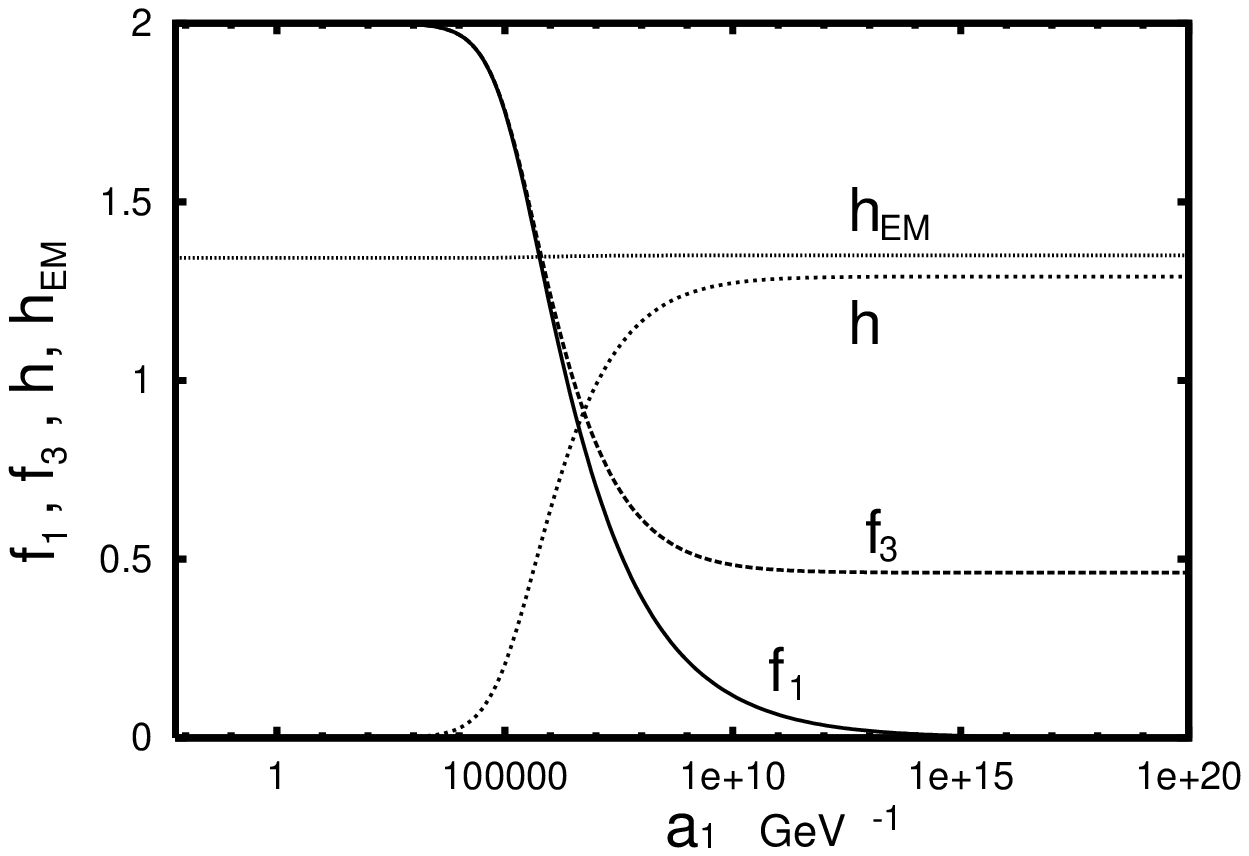}}   
\caption{The evolution of $f_1, f_3, h, h_{\rm EM}$ when
the configuration started from the local minimum of the potential 
when $\beta < \sqrt{3\alpha}$.
In this example, $\Lambda= 1.0 \times 10^{5}$ GeV$^2$ and $a_{1}(t_0)= 
a_{3}(t_{0}) =6 \times 10^{-3} {\rm GeV}^{-1}$.  
}
\label{fh-evolution1-2}
\end{figure}

\noindent
Figures \ref{fh-evolution1} and \ref{fh-evolution1-2} 
correspond to Case IIa and IIb respectively.   
The evolution of $f_1, f_3, h$ and
$h_{\rm EM}$ is depicted as a function of $a_1(t)$.  
In fig.\  \ref{fh-evolution1}  the initial  fields
are located at the local minimum of the potential, 
$(f_1, f_3, h, v)= (1.985, 1.985, 0.03307, 158.6 \,{\rm GeV})$.  $f_1$ and 
$h_{\rm Z} \propto f_3-g'h$ approach zero around $a=10^3$GeV$^{-1}$, 
but $f_3$ and $h$  remain non-vanishing.
$h_{\rm EM}$ asymptotically approaches a non-vanishing value in accordance 
with our previous argument.  The magnetic
field $h_{\rm EM}/a_1^2$ thus produced is  decreasing only by 
the factor $a_{1}^{2}$.
In Figure \ref{fh-evolution1-2}, the fields initially start 
a tiny bit off the local minimum  $(f_1, f_3, h, v)= (2.0, 2.0, 0,
0.1,{\rm GeV})$.   The corresponding $\beta  < \sqrt{3\alpha}$  in
(\ref{condition3}).
 
We observe in this case that $f_{1}$ and $f_{2}$ stay at the minimum 
for some time and then decrease. $h$ also stays at $0$ for a while 
and start to move towards the asymptotic value. Here again, we observe 
that $h_{\rm EM}$  approaches non-vanishing value.

\begin{figure}[tbp]\centering
\leavevmode 
\mbox{
\epsfxsize=8cm 
\epsfbox{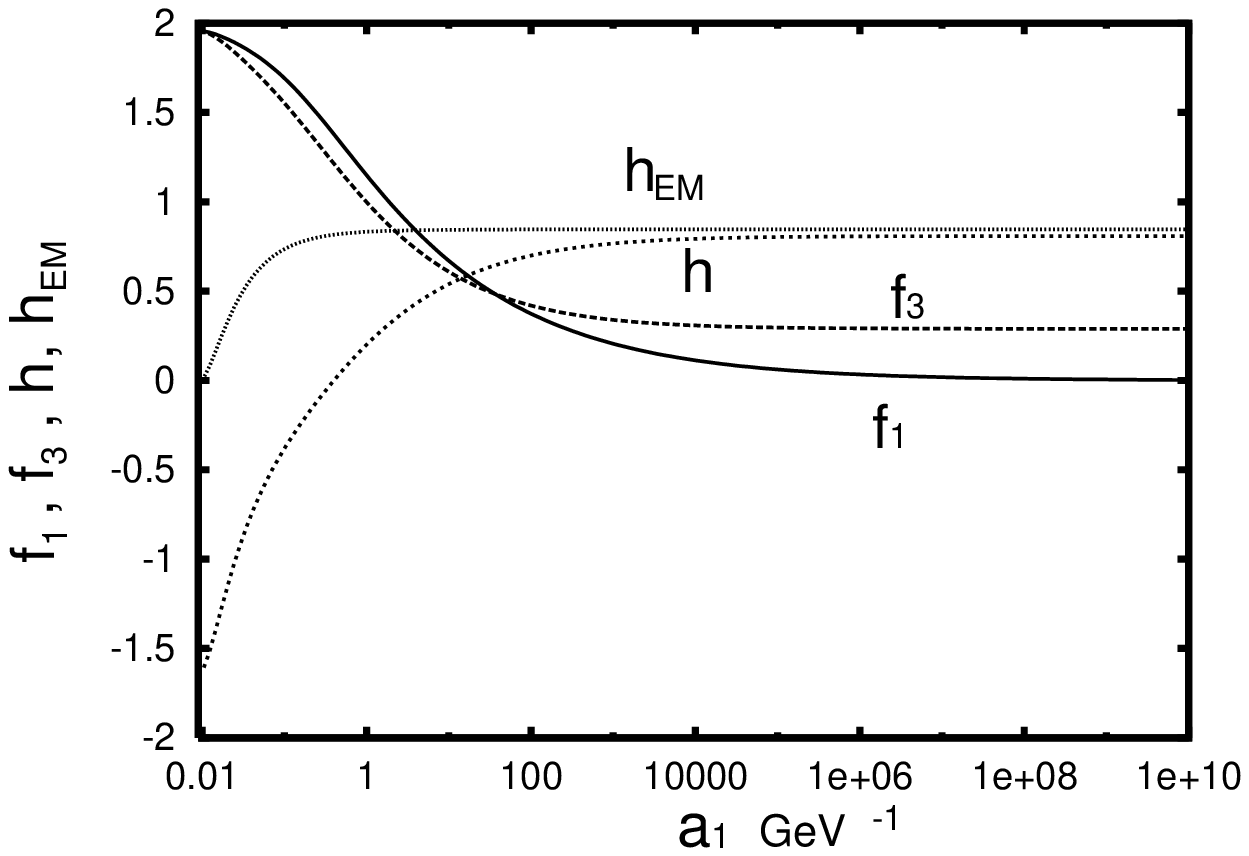}}   
\caption{The evolution of $f_1, f_3, h, h_{\rm EM}$ when
the configuration started off the minimum of the potential.
We put
$\Lambda=1.0 \times 10^5 {\rm GeV}^2$ and $a_{1}(t_0)=a_{3}(t_{0})=
9.0 \times10^{-3}{\rm GeV}^{-1}$}
\label{fh-evolution2}
\leavevmode 
\mbox{
\epsfxsize=8cm 
\epsfbox{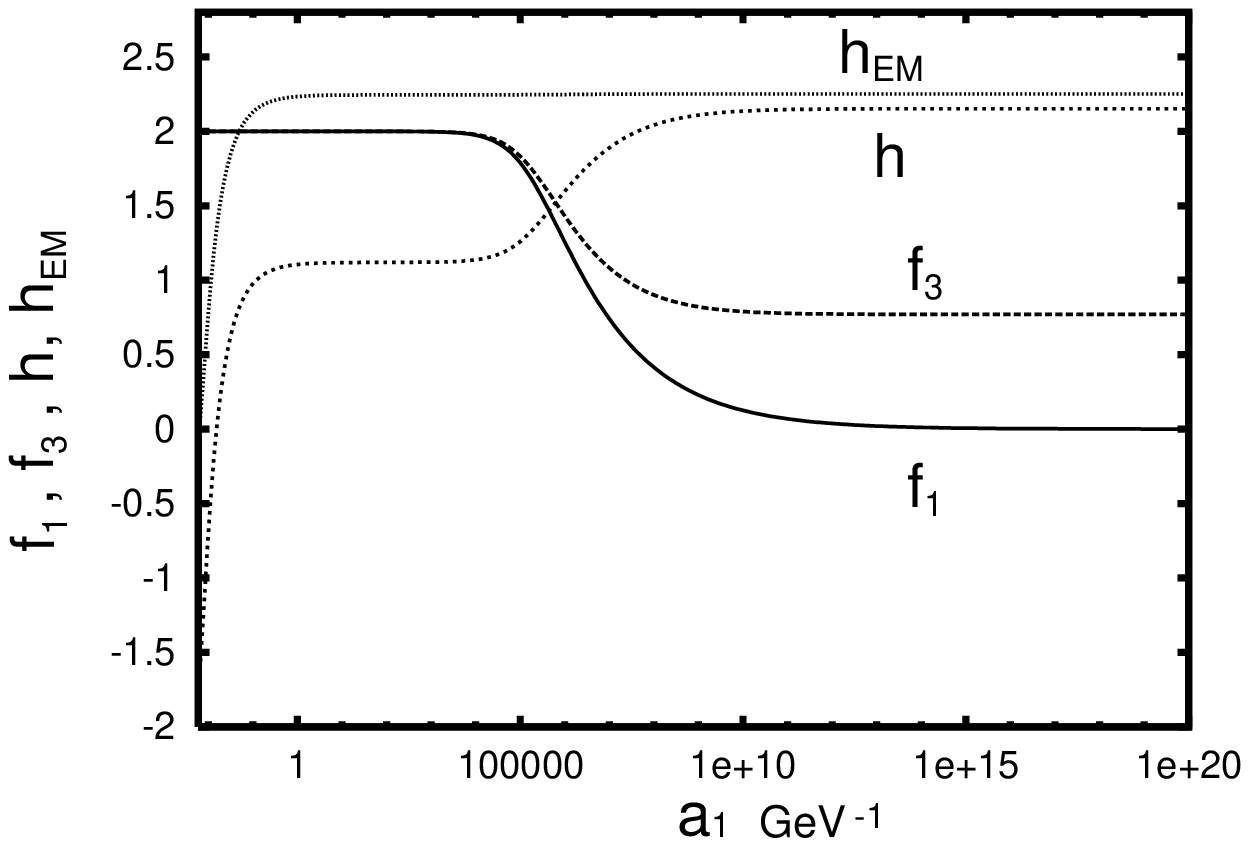}}   
\caption{The evolution of $f_1, f_3, h, h_{\rm EM}$ when
the configuration started off the minimum of the potential when $\beta < \sqrt{3\alpha}$.
We put
$\Lambda=1.0 \times 10^5 {\rm GeV}^2$ and $a_{1}(t_0)=a_{3}(t_{0})=
6.0 \times10^{-3}{\rm GeV}^{-1}$}
\label{fh-evolution2-2}
\end{figure}

One may perhaps wonder what would happen if we  start 
with a vanishing magnetic field $h_{\rm EM}(t_{0})=0$. 
It is of  interest to start not exactly from the 
local minimum but  away from it, thereby adjusting to 
$h_{\rm EM}(t_{0})=0$.     
An example is displayed in Fig.\ \ref{fh-evolution2}.
the initial conditions are 
$(f_1, f_3, h, h_{\rm EM}, v)= (1.958, 1.958, -1.644, -0.009838, 129.8 \,{\rm GeV})$ at $t_0$.
Note that the fields immediately start to roll down.  
$f_1$ and $h_{\rm Z}$ quickly approach zero.  However, $h_{\rm EM}$, which
started with a vanishing value, quickly gains a nonvanishing value.  The final
value depends on the initial condition, but the fact that $h_{\rm EM}$
approaches a nonvanishing value does not. Fig.\ \ref{fh-evolution2} 
illustrates this remarkable fact. 

\begin{figure}[tbp]\centering
\leavevmode 
\mbox{
\epsfxsize=8cm 
\epsfbox{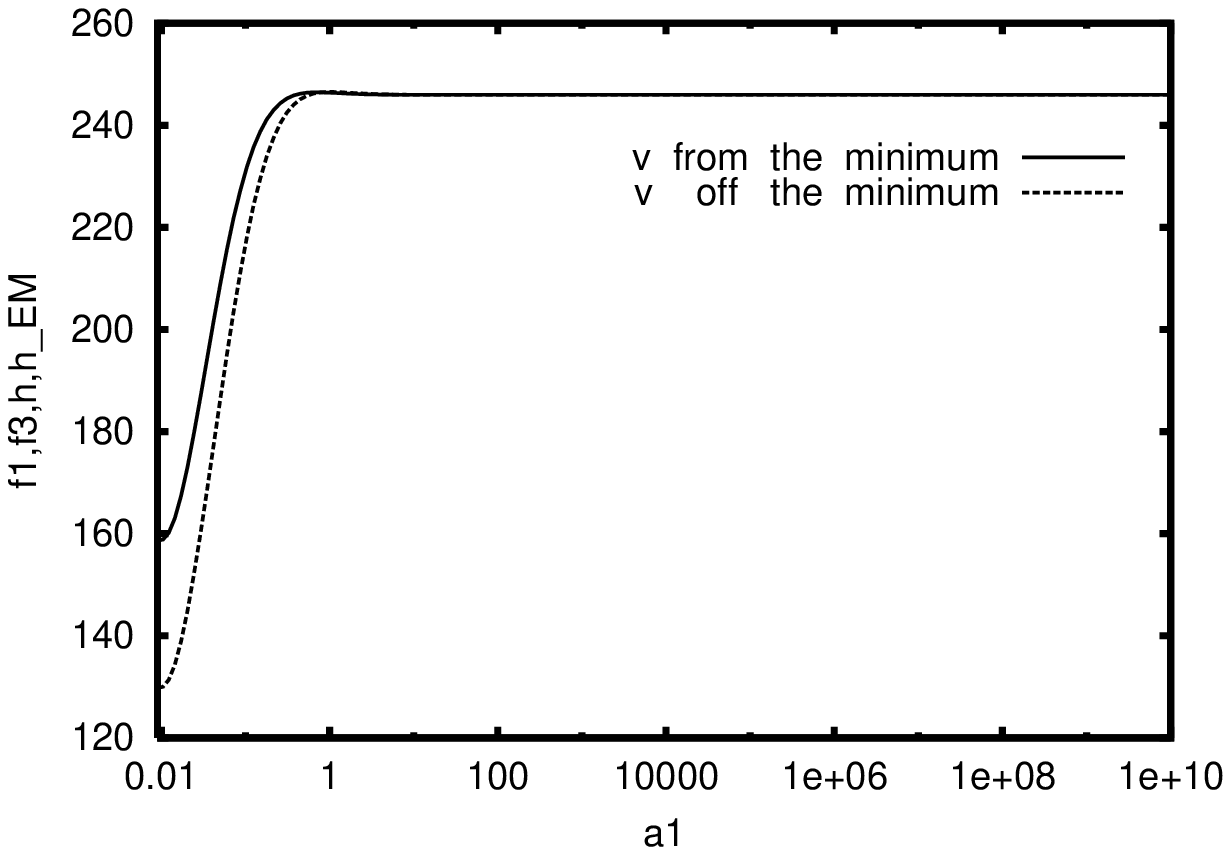}}   
\caption{The evolution of $v$ when the configuration started from and 
away from the local minimum of the potential. The situation for ``from''
 and ``off'' the minimum correspond to those of Figure 6 and Figure 8
respectively.}
\label{v-evolution}
\leavevmode 
\mbox{
\epsfxsize=8cm 
\epsfbox{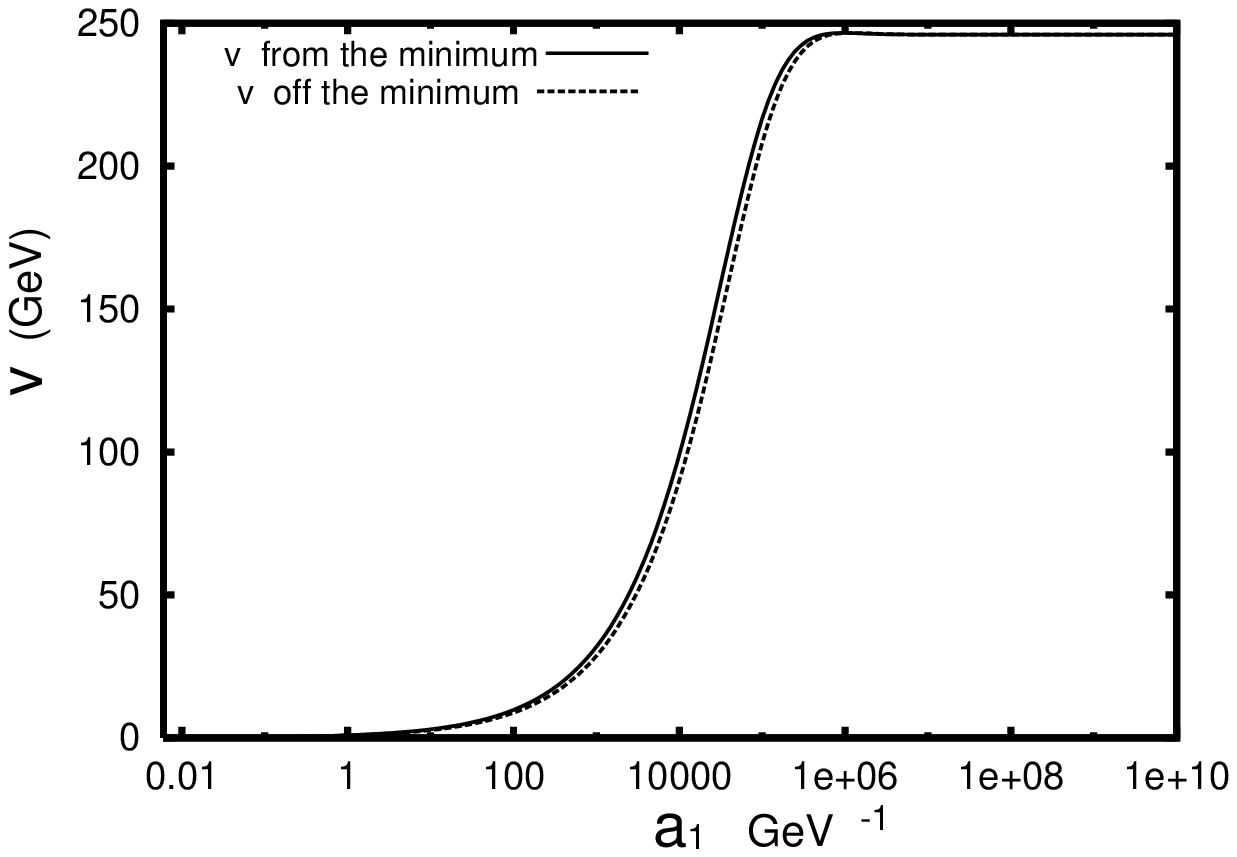}}   
\caption{The evolution of $v$ when the configuration started from and 
away from the local minimum of the potential. The situation for ``from''
 and ``off'' the minimum when $\beta < \sqrt{3\alpha}$ correspond to 
those of Figure 7 and Figure 9 respectively.}
\label{v-evolution2}
\end{figure}

The behavior of the Higgs field $v$ is depicted in Fig.\
\ref{v-evolution}.  It approaches
the value $v_0$ at the global minimum, much in the same way as the gauge
fields $f_1$ and $h_Z$.  It leads to  symmetry breaking as in the usual
case.

\bigskip

\leftline{\bf Case III. $\Lambda \ll (g v_0)^2$ ,  
  $\rho \ll  (10^{11} \, {\rm GeV})^{4}$}

In this case the potential $V$ in (\ref{potential4}) does not have
a local minimum.  
Nevertheless it is of great interest to ask what
would happen if the universe, at one instant $t_0$, assumes nonvanishing 
values for $f_1, f_3$ and $h$.  The size $a_j(t_0)$ of the universe
has to be very large to be consistent with the Einstein equations,
which in turn implies that resultant field strengths of the gauge
fields are negligibly small.  

\begin{figure}[tbh]\centering
\leavevmode 
\mbox{
\epsfxsize=10cm 
\epsfbox{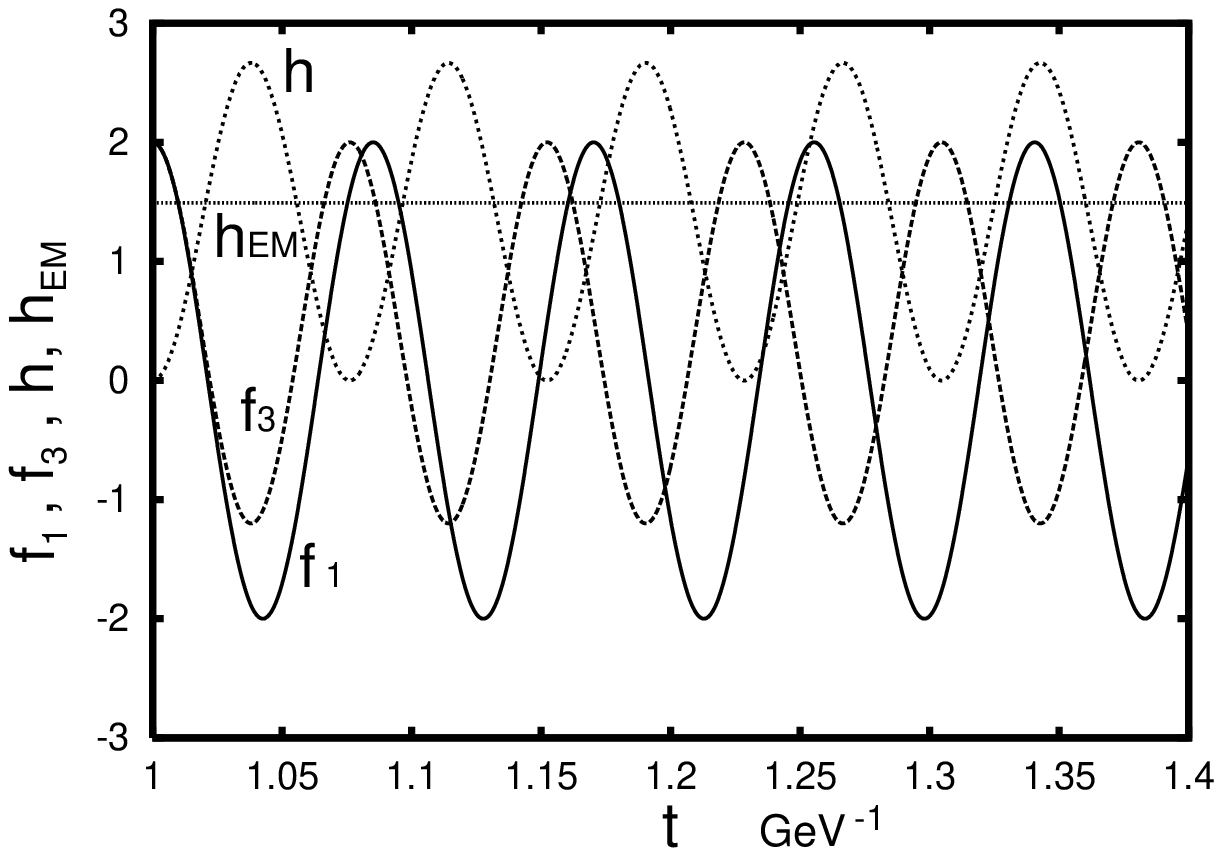}}   
\caption{Time evolution of $f_1, f_3, h$ $h$ and $h_{\rm EM}$, 
when $\Lambda = 1.0 \times 10^{-29} {\rm GeV}^{2}$, $a_{1}(t_0)= 
a_{3}(t_{0}) = 4.5\times 10^{14} {\rm GeV}^{-1}$  }
\label{field-fig3}
\end{figure}

As an example we set $\Lambda=1.0 \times 10^{-29} {\rm GeV}^{2}$
corresponding to $\rho=\Lambda/8\pi G= v_0^4$.   (??)
We pick up an initial configuration with $a_j=4.5\times 10^{14}\,$GeV$^{-1}$,
$f_1=f_3=2$, $h=0$, and $v=v_0$.   The evolution is
plotted in Fig.\ \ref{field-fig3}.  
The gauge fields $f_{1}$, $f_{3}$, and $h$ show oscillatory behavior.
They  vary in the time scale of 0.05 ${\rm GeV}^{-1}$.
As $a_j$ is very large,
$V_4$ is totally irrelevant.  $V_2$  quadratically depends on 
$f_1$ and $h$.  As $a_j$ varies very slowly, its $t$-dependence may be
ignored.   $f_1$ and $h_Z$, then, 
exhibit harmonic oscillation with frequency $\omega = {1\over2} gv_0$
and ${1\over 2} (g^2 +g'^2)^{1/2} v_0$, respectively.   
$h_{\rm EM}$, on the other hand, remains constant.

\sxn{Generation of  electromagnetic field}

One interesting aspect of the cosmological evolution of the 
nontrivial topological configuration is that electromagnetic
fields are produced over a substantial period of the expansion of 
the universe.  In this paper we have examined only the bosonic sector of the
electroweak theory, ignoring quarks and leptons. It is expected that
once dynamics of quarks and leptons are included, the presence of 
large electromagnetic fields triggers pair creation of fermions, thus
affecting the subsequent evolution of the universe.  In this section
we would like to see how large electromagnetic fields are, and
how they depend on the parameters of the theory.

\begin{figure}[tbh]\centering
\leavevmode 
\mbox{
\epsfxsize=10cm 
\epsfbox{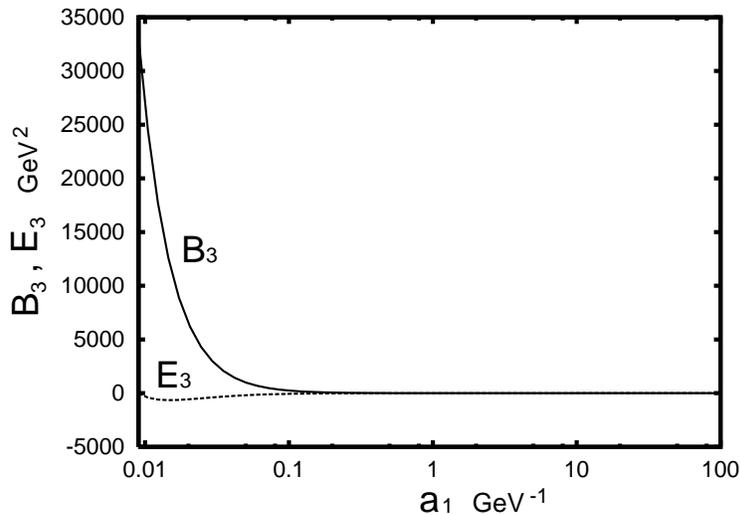}}   
\caption{Electromagnetic fields when the configuration starts from  the
local minimum.  $\Lambda=1.0 \times 10^5 {\rm GeV}^2$ and
$a_{1}(t_0)=a_{3}(t_{0})= 9.0 \times10^{-3} {\rm GeV}^{-1}$ }
\label{EMlocal}
\end{figure}

We recall that the electric and magnetic fields are given by
$E_3 = \dot h_{\rm EM}/a_3$ and $B_3 = 2h_{\rm EM}/a_1^2$.
As we have seen, generated $h_{\rm EM}$ is typically ${\cal O}(1)$ so 
that the electromagnetic
fields become relevant only when $a_j$'s are sufficiently small.
In fig.\ \ref{EMlocal}  the evolution of the electromagnetic fields 
is displayed in which the field configuration starts from the local
minimum. (We take the same initial condition as in fig.\
\ref{fh-evolution1}.) 
  One sees that the magnetic field $B_{3}$ persists to exist for 
considerable time. \cite{Barrow}

\begin{figure}[htb]\centering
\leavevmode 
\mbox{
\epsfxsize=10cm 
\epsfbox{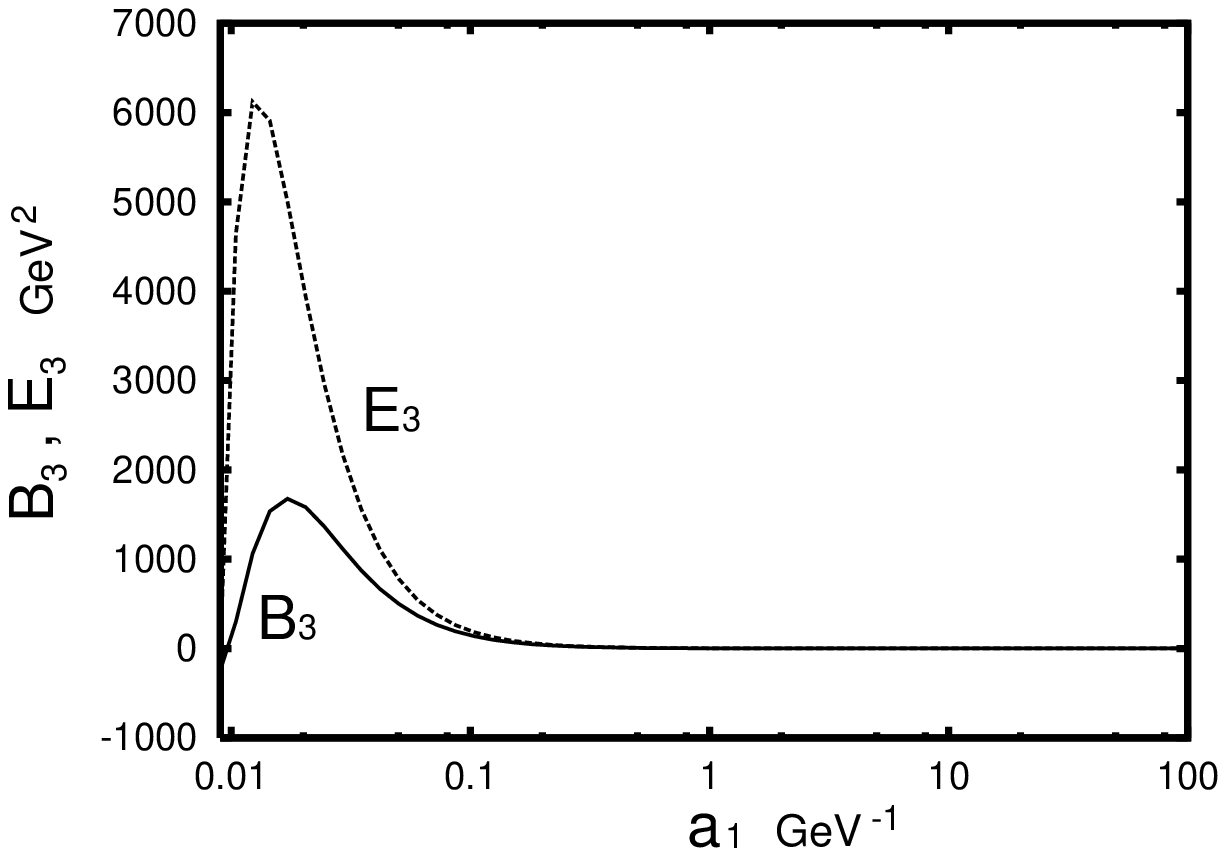}}   
\caption{Electromagnetic fields when the configuration starts  away from 
the minimum.  $\Lambda=1.0 \times 10^5 {\rm GeV}^2$ and
$a_{1}(t_0)=a_{3}(t_{0})= 9.0 \times10^{-3} {\rm GeV}^{-1}$}
\label{EMgeneral}
\end{figure}

\begin{figure}[htb]\centering
\leavevmode
\mbox{
\epsfxsize=8cm 
\epsfbox{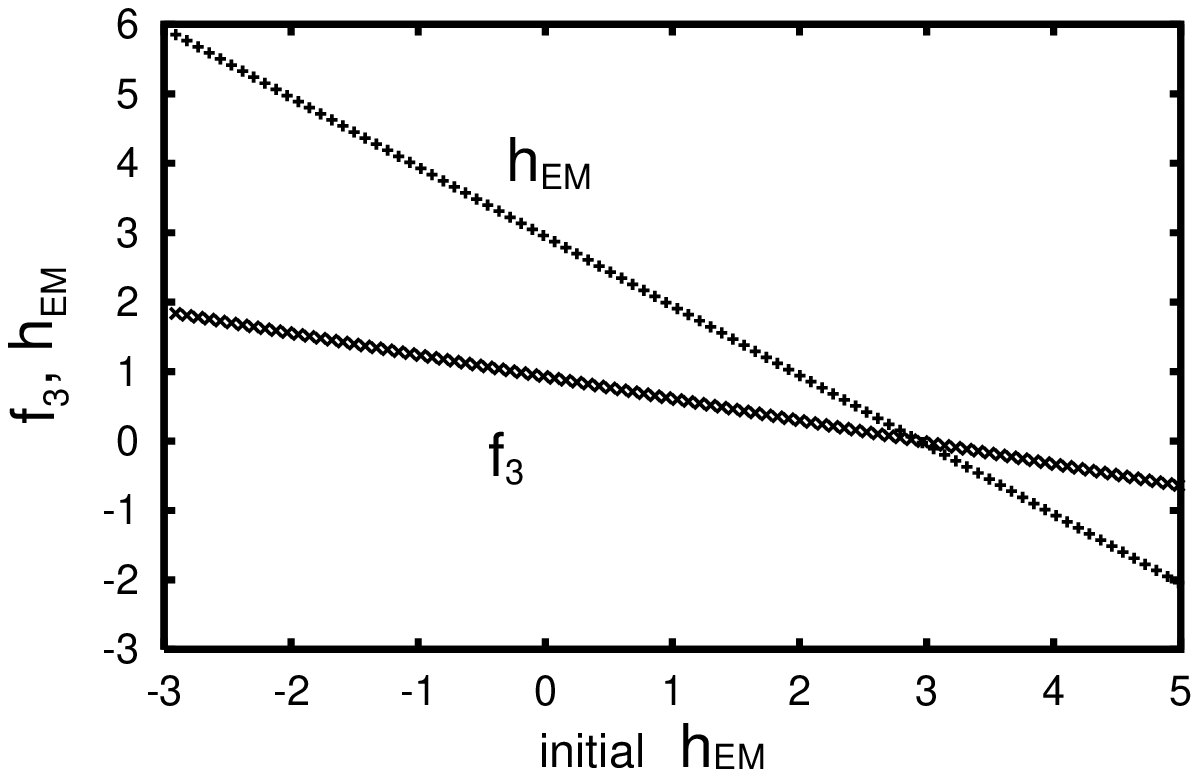}}   
\caption{The dependence of the generated electromagnetic
field $h_{\rm EM}$ and $f_3$ on the initial $h_{\rm EM}$ imposed is
plotted.  $\Lambda = 1 \times 10^6$ GeV$^2$ and $a_j(t_0)=
\sqrt{3/\Lambda}$.  The linear dependence is observed.
}
\label{in-out}
\end{figure}

One may wonder if the magnetic field just found is a result of a
special initial condition chosen, and
if it can be generated even with a more general initial condition.
To have a more careful look at this point, another example 
of the evolution is displayed  in fig.\ \ref{EMgeneral}. Here we have
adjusted  the initial magnetic field as $B_{3}=0$, choosing the starting
point  away from the local minimum. (We  take the same initial  condition 
as in Fig.\ \ref{fh-evolution2}.)  We clearly see that the
magnetic field is indeed generated for substantially long period.  
In Fig.  \ref{in-out} the dependence on the initial $h(t_0)$ is plotted with 
other parameters fixed.  Amusingly the linear dependence is observed.

\begin{figure}[tbh]\centering
\leavevmode 
\mbox{
\epsfxsize=10cm 
\epsfbox{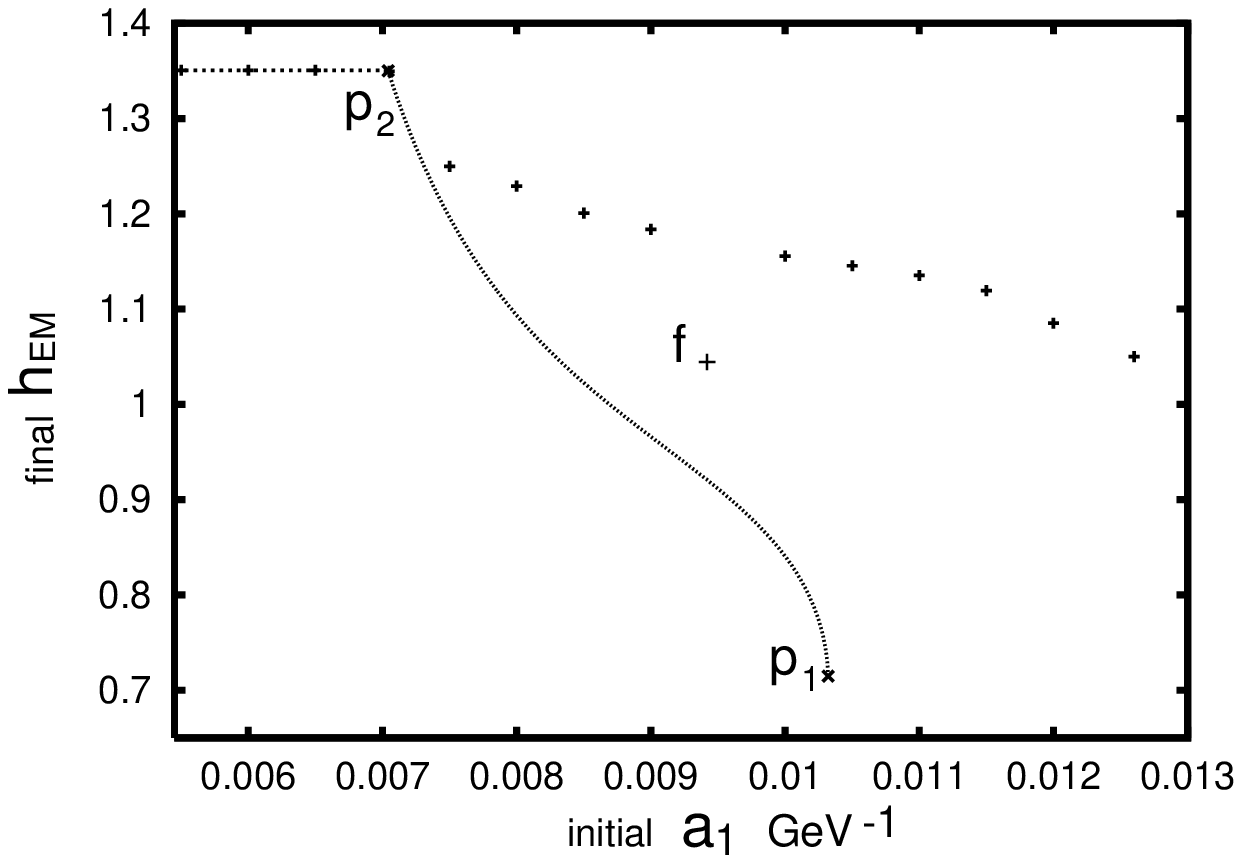}}   
\caption{$a_{1}$-dependence of final $h_{\rm EM}$. 
$\Lambda =1.0\times 10^5{\rm GeV^2}$.  Points + and the dotted line correspond
to configurations starting from the local 
minimum of $SU(2)_L\times U(1)_Y$ potential 
and $(f,v)=(f_+, v_+)$ in (\ref{fplusminus}) and (\ref{condition2}),
respectively.  Points $p_1, p_2$ and the line $f_+$ correspond to 
those in figs.\  2 and 3.}
\label{a-dependence}
\end{figure}

The final value of $h_{\rm EM}$ depends  on the initial condition.  
In Fig.\ \ref{a-dependence}  the dependence of $h_{\rm EM}$ upon 
$a_j(t_0)$ or $\beta(t_0)$ is plotted with   $\Lambda=1.0\times
10^5{\rm GeV^2}$ given. Here  the initial  configuration is set at  the
local minimum of the  potential. There  exists little
$a_{1}$-dependence for small $\beta$, but has weak dependence near
$\beta_c$.    
We can safely conclude that  the existence of the magnetic
field for a substantial time  during in the early universe  is a generic
and quite probable phenomenon.  

Another interesting question arises about the $\Lambda$-dependence
on the final value of $h_{\rm EM}$.
In varying $\Lambda$, we always take $a_j^2 = 3/\Lambda$ initially, 
 the field configuration residing  at the local minimum. 
It turns out the asymptotic values of the fields do not depend
on $\Lambda$, though the transition scale shows the dependence 
(\ref{transition}).

\sxn{Summary}

In the present paper we have explored the interplay between the
$SU(2)_{L} \times U(1)_{Y}$  electroweak interactions and gravity, 
especially in the context of the expanding universe.  We have unveiled 
that in the Einstein-electroweak theory there exists a
nontrivial topological configuration of the Higgs and gauge fields 
which corresponds to  a local minimum of the potential in the field space.
We have looked into the time  evolution and fate of such  a nontrivial
configuration to discover an interesting conspiracy by
the gravity-gauge-Higgs system.
Even if the gauge-Higgs system is initially endowed with  non-trivial
topology, the system   cannot maintain the topology perpetually.   As the
universe expands,   the potential of the  gauge-Higgs system undergoes a
change, the  barrier separating  the non-trivial and trivial
configurations thereby disappearing.  The gauge and Higgs fields  start to
roll down the hill in the potential toward a configuration with a lower
energy.  However,  if the universe expands fast enough, say, 
driven by an effective  cosmological constant, then the fields can never
reach the global minimum of the potential.   The electromagnetic field
$h_{\rm EM}$  survives in the evolution.  It is  generated  for a wide
range of parameters. The space of the resultant universe is a deformed
$S^3$ which is homogeneous but anisotropic.

It is of great interest to apply  our findings in the actual history of
the universe. In the standard scenario of the early universe,
the temperature  effect which we have neglected throughout is  
important.  It modifies the shape of the potential 
as well as the time-evolution.   What we have in mind as one possible
scenario is an era preceding the hot universe which continuously  evolves
to the current universe by radiation or matter dominance.   We suppose
that at one instant the universe was very small and cool, and the gauge
and Higgs fields assumed a nontrivial configuration. Driven by an
effective cosmological constant the
universe underwent inflation.   In a substantial  period in the expansion
sizable electromagnetic fields were generated. Eventually the inflation
stopped and the universe was reheated to the temperature
about $(\Lambda/8\pi G)^{1/4}$.  The universe   continued to expand by
the radiation dominance since then.

It would be very interesting to investigate  consequences of 
strong electromagnetic fields thus generated.   
In such strong background of electric and magnetic fields, there could be
phenomena  such as pair creation of fermions and might affect the relic
abundance  of various elements.  All these problems are left for  our
future work.

\vskip .5cm

\leftline{\bf Acknowledgments}

This work was supported in part by   by Scientific Grants 
from the Ministry of Education and Science, Grant Number 13135215 (Y.H.\ 
and T.K.) and Grant Number 13640284 (Y.H.).  One of the authors (Y.H.)
would like to thank Yoichiro Nambu for valuable comments in the early
stage of the investigation and  Naoki Sasakura for pointing out an error
in the preliminary version of the work, respectively.

\vskip 1cm

\axn{Geometry of  $S^3$ and deformed $S^3$ }

Let us summarize the Riemann, Ricci and scalar curvatures 
of the deformed sphere $S^3$, whose   metric is given by 
\begin{eqnarray}
& &ds^2=-N(t)^{2}dt^{2}
  + \sum_{j=1}^3 a_j(t)^{2} \sigma^j   \otimes \sigma^j
  =\eta_{a b} e^a \otimes  e^b
\label{general-metric}
\end{eqnarray}
where use has been made of the local Lorentz metric 
$\eta_{a b}={\rm diag} (-1,1,1,1)$  and tetrad bases 
\begin{eqnarray}
e ^{0}= N(t)dt,\qquad \qquad e^{i}=a_{i}(t)\sigma^{i}, \qquad (i=1,2,3).
\end{eqnarray}
In the main body of the present paper we have set 
\begin{eqnarray}
a_1(t)=a_2(t)=a_3(t)=a(t) 
\end{eqnarray}
for $\theta _{W}=0$ case and
\begin{eqnarray} 
a_1(t)=a_2(t) 
\end{eqnarray}
for $\theta _{W}\neq 0$ case. Here, however, we keep 
our metric (\ref{general-metric}) as general as possible by putting the 
three scale factors $a_{i}(t)$ 's ($i=1,2,3)$ on an equal footing.

We impose conditions for vanishing torsion 
\begin{eqnarray}
 de^a + {\omega^a}_b \wedge e^a =0, \qquad (a, b =0,1,2,3)
\end{eqnarray}
and express the connections   
$\omega_{ab}= -\omega_{ba}$
in terms of the tetrads. 
Straightforward calculations lead
\begin{eqnarray}
\omega_{0i}= - \frac{\dot{a_i}}{a_iN}e^i, 
\qquad 
\omega_{ij} =\epsilon_{ijk} e^{k} \tilde \omega ^{k}
\label{eq:setsuzoku}
\end{eqnarray}
Here we have introduced  notation 
\begin{eqnarray}
\tilde \omega ^{k}=\frac{a_{\ell}}{a_{k}a_{m}}+
\frac{a_{m}}{a_{k}a_{\ell}}-\frac{a_{k}}{a_{\ell}a_{m}}
\end{eqnarray}
and indices $(k, \ell , m)$ are cyclic permutations of (1, 2, 3).

Curvature 2-form is defined as
\begin{eqnarray}
{\mathcal{R}^a}_b={d\omega^a}_b +{\omega^a}_c\wedge {\omega^c}_b
=\frac{1}{2}{R^a}_{bcd}e^c\wedge e^d
\label{eq:kyokuritsu}
\end{eqnarray}
By putting (\ref{eq:setsuzoku}) into (\ref{eq:kyokuritsu}), 
we obtain curvature 2-forms as follows:
\begin{eqnarray}
{\mathcal R}_{0 i}=-{\mathcal R}_{i 0}=- \frac{1}{Na_i}\frac{d}{dt}
\left (\frac{\dot{a_i}}{N} \right )e^0\wedge e^i
-\frac{1}{N}\left ( \frac{2\dot{a_i}}{a_ja_k}- \frac{\dot{a_j}}{a_j}
\tilde \omega ^{k}-\frac{\dot{a_k}}{a_k}\tilde \omega ^{j}
\right ) e^j \wedge e^k , 
\label{eq:0i}
\\
{\mathcal R}_{i j}=-{\mathcal R}_{j i}=\frac{1}{Na_k}\frac{d}{dt}\left (
a_k \tilde \omega ^{k} \right ) e^0 \wedge e^k + 
\left (  \frac{\dot{a_i}\dot{a_j}}{N^2 a_i a_j} 
+ \frac{2a_k}{a_ia_j}  \tilde \omega ^k  
-  \tilde \omega ^{i} \tilde \omega ^{j}  \right ) e^i \wedge e^j .
\label{eq:ij}
\end{eqnarray}
The indices $(i, j, k)$ are cyclic permutations of (1, 2, 3) and repeated 
indices are not summed over in (\ref{eq:0i}) or (\ref{eq:ij}), either.
Each component of the Riemann tensors can be easily read off by comparing 
(\ref{eq:0i}) and (\ref{eq:ij}) with (\ref{eq:kyokuritsu}). 
\ignore{
For completeness we list them up:  
\begin{eqnarray}
R_{0i 0 i}&=&-\frac{1}{Na_i}\frac{d}{dt}\left ( \frac{\dot{a_i}}{N}\right )
\\
R_{0ijk}&=& -\frac{1}{N}\left ( \frac{2\dot{a_i}}{a_j a_k}
- \frac{\dot{a_j}}{a_j}  \tilde \omega ^{k} 
- \frac{\dot{a_k}}{a_k}\tilde \omega ^{j} \right ) 
\\
R_{ij0k}&=&\frac{1}{N a_k}\frac{d}{dt}\left (a_k \tilde \omega ^{k} \right )
=R_{0kij}
\\
R_{ijij}&=&\frac{\dot{a_i}\dot{a_j}}{N^2a_ia_j} 
+\frac{2a_k}{a_ia_j}\tilde \omega ^{k}-\tilde \omega ^{i}\tilde \omega ^{j}
\end{eqnarray}
The indices $(i,j,k)$ above are again cyclic permutations of (1,2,3). 
The unlisted components are either vanishing or obtained from above by 
symmetry properties $R_{abcd}=R_{cdab}=-R_{bacd}=-R_{abdc}$.
}

The  Ricci tensor is non-vanishing only for diagonal components:
\begin{eqnarray}
R_{00}&=&-\frac{1}{N}\left \{\frac{1}{a_1}\frac{d}{dt}\left (
\frac{\dot{a_1}}{N}\right )+ \frac{1}{a_2}\frac{d}{dt}\left ( 
\frac{\dot{a_2}}{N}\right )+ \frac{1}{a_3}\frac{d}{dt}\left ( 
\frac{\dot{a_3}}{N}\right ) \right \}
\\
R_{ii}&=&\frac{1}{Na_i}\frac{d}{dt}\left (\frac{\dot{a_i}}{N}\right )
+\frac{4}{(a_i)^2}+\frac{\dot a_{i}}{Na_{i}}\left (
\frac{\dot a_{j}}{Na_{j}}+\frac{\dot a_{k}}{Na_{k}}
\right )
\nonumber \\
& & +2 \left \{
\left ( \frac{a_{i}}{a_{j}a_{k} }  \right )^2 -
\left ( \frac{a_{j}}{a_{k}a_{i} }  \right )^2 -
\left ( \frac{a_{k}}{a_{i}a_{j} }  \right )^2 
\right \}
\end{eqnarray}
As before the indices $(i, j, k)$ are cyclic permutations of (1, 2, 3).
Notice that the Riemann and Ricci tensors in the tetrads depend on
time $t$, but not on spatial coordinates.  The spacetime defined by
metric  (\ref{general-metric}) is spatially homogeneous, but
anisotropic. The scalar curvature is given by 
\begin{eqnarray}
R&=&=-R_{00}+R_{11}+R_{22}+R_{33}
\nonumber \\
&=&\sum _{i=1}^{3} \left \{
\frac{2}{N a_i} \frac{d}{dt}\left ( \frac{\dot{a_i}}{N} \right )
+ \frac{4}{(a_{i})^{2}}\right \}
+\frac{2}{N^2}\left (\frac{\dot{a_1}\dot{a_2}}{a_1a_2}
                    +\frac{\dot{a_2}\dot{a_3}}{a_2a_3}
                    +\frac{\dot{a_3}\dot{a_1}}{a_3a_1}
                    \right )
\nonumber \\
& & - 2 \left \{
 \left (\frac{a_{1}}{a_{2}a_{3}}\right )^{2}
+\left (\frac{a_{2}}{a_{3}a_{1}}\right )^{2}
+\left (\frac{a_{3}}{a_{1}a_{2}}\right )^{2}
\right \}
\end{eqnarray}

\vskip .5cm

\def\jnl#1#2#3#4{{#1}{\bf #2} (#4) #3}

\def\Zphys{{\em Z.\ Phys.} }
\def\jssc{{\em J.\ Solid State Chem.\ }}
\def\jpsJ{{\em J.\ Phys.\ Soc.\ Japan }}
\def\ptps{{\em Prog.\ Theoret.\ Phys.\ Suppl.\ }}
\def\PTP{{\em Prog.\ Theoret.\ Phys.\  }}

\def\JMP{{\em J. Math.\ Phys.} }
\def\NPB{{\em Nucl.\ Phys.} B}
\def\NP{{\em Nucl.\ Phys.} }
\def\PLB{{\em Phys.\ Lett.} B}
\def\PL{{\em Phys.\ Lett.} }
\def\PRL{\em Phys.\ Rev.\ Lett. }
\def\PRB{{\em Phys.\ Rev.} B}
\def\PRD{{\em Phys.\ Rev.} D}
\def\PRe{{\em Phys.\ Rep.} }
\def\AP{{\em Ann.\ Phys.\ (N.Y.)} }
\def\RMP{{\
em Rev.\ Mod.\ Phys.} }
\def\ZPC{{\em Z.\ Phys.} C}
\def\SCI{\em Science}
\def\CMP{\em Comm.\ Math.\ Phys. }
\def\MPLA{{\em Mod.\ Phys.\ Lett.} A}
\def\IJMPB{{\em Int.\ J.\ Mod.\ Phys.} B}
\def\PR{{\em Phys.\ Rev.} }
\def\cmp{{\em Com.\ Math.\ Phys.}}
\def\JPA{{\em J.\  Phys.} A}
\def\CQG{\em Class.\ Quant.\ Grav. }
\def\ATMP{{\em Adv.\ Theoret.\ Math.\ Phys.} }
\def\ibid{{\em ibid.} }

\leftline{\bf References}

\renewenvironment{thebibliography}[1]
        {\begin{list}{[$\,$\arabic{enumi}$\,$]}  
        {\usecounter{enumi}\setlength{\parsep}{0pt}
         \setlength{\itemsep}{0pt}  \renewcommand{\baselinestretch}{1.2}
         \settowidth
        {\labelwidth}{#1 ~ ~}\sloppy}}{\end{list}}

\end{document}